\documentclass[apj]{emulateapj}

\usepackage{graphicx,times}
\usepackage{subfigure}
\newcommand{\be}{\begin{equation}}
\usepackage{threeparttable}
\usepackage{booktabs}
\newcommand{\ee}{\end{equation}}
\newcommand{\bea}{\begin{eqnarray}}
\newcommand{\eea}{\end{eqnarray}}

\usepackage{enumerate}
\usepackage{amsmath}
\usepackage{cases}
\usepackage{longtable}
\usepackage{hyperref}
\usepackage{epstopdf}
\usepackage{amsmath,bm}
\usepackage{amssymb}
\usepackage{natbib}
\usepackage{morefloats}
\usepackage{multirow}
\usepackage{array}
\usepackage{verbatim}

\begin{document}

\title{Resonance-broadened transit time damping of particles in MHD turbulence}

\author{Siyao Xu\altaffilmark{1,2} and Alex Lazarian\altaffilmark{1} }

\altaffiltext{1}{Department of Astronomy, University of Wisconsin, 475 North Charter Street, Madison, WI 53706, USA; 
sxu93@wisc.edu,
lazarian@astro.wisc.edu}
\altaffiltext{2}{Hubble Fellow}

\begin{abstract}

As a fundamental astrophysical process, 
the scattering of particles by turbulent magnetic fields has its physical foundation laid by the magnetohydrodynamic (MHD) turbulence theory. 
In the framework of the modern theory of MHD turbulence, 
we derive a generalized broadened resonance function by taking into account both the 
magnetic fluctuations and nonlinear decorrelation of turbulent magnetic fields arising in MHD turbulence, 
and we specify the energy range of particles for the dominance of different broadening mechanisms.  
The broadened resonance allows for scattering of particles beyond the energy threshold of the linear resonance. 
By analytically determining the pitch-angle diffusion coefficients for transit time damping (TTD) with slow and fast modes, 
we demonstrate that the turbulence anisotropy of slow modes suppresses their scattering efficiency. 
Furthermore, we quantify the dependence of the relative importance between slow and fast modes in TTD scattering on 
(i) particle energy, 
(ii) plasma $\beta$ (the ratio of gas pressure to magnetic pressure), 
and (iii) damping of MHD turbulence,
and we also provide the parameter space for the dominance of slow modes. 
To exemplify its applications, we find that among typical partially ionized interstellar phases, in the warm neutral medium slow and fast modes 
have comparable efficiencies in TTD scattering of cosmic rays.  
For low-energy particles, e.g., sub-Alfv\'{e}nic charged grains, 
we show that slow modes always dominate TTD scattering. 

\end{abstract}

\keywords{turbulence - magnetic fields - cosmic rays }

\section{Introduction}

The scattering and acceleration of particles are the central ingredients of space physics and astrophysics
\citep{Longairbook}.
They play crucial roles in many phenomena in the 
heliosphere, the interstellar medium (ISM), and the intracluster medium
\citep{Schlickeiser02,BruLaz11}. 
The scattering of particles is usually described by the pitch-angle diffusion coefficient, 
which is directly related to the parallel spatial diffusion and acceleration of particles.

The pitch-angle diffusion coefficient is conventionally calculated using the quasi-linear theory (QLT)
\citep{Jokipii1966}, 
under the assumption of unperturbed orbits of particles in weakly turbulent plasmas. 
As a result, earlier studies of particle transport and acceleration were restricted to the case of 
slab model of turbulence
\citep{Sch94}.
However, the slab model of turbulence is not supported by 3D magnetohydrodynamic (MHD) simulations 
\citep{CV00,MG01,CL03}, 
and the diffusion behavior of particles in such turbulence geometry also does not reconcile with observations
\citep{Palmer:1982,Wan93}.
Besides, the QLT is not applicable to studying 
the $90^\circ$ scattering 
\citep{Jones:1973}
and perpendicular diffusion  
\citep{Kota_Jok2000}
of particles.

To study the interaction of particles with the turbulent magnetic fields permeating the interstellar and interplanetary media, 
a realistic description of MHD turbulence is the key element. 
The synthetic turbulence models, e.g., slab/2D composite model
\citep{Mat90},
a superposition of plane waves
\citep{Giacalone_Jok1999},
have been formulated to mimic some of the 
fundamental characteristics of MHD turbulence, such as the anisotropy and the spectral form. 
Nevertheless, these phenomenological models cannot capture the nature of MHD turbulence
and are in contradiction with direct numerical simulations of 3D MHD turbulence
\citep{CV00,MG01,CLV_incomp, CL03, KL12}.
These simulations instead support the physically motivated turbulence model developed by 
\citet{GS95} (hereafter GS95). 
The GS95 model provides a self-consistent interpretation of 
the anisotropic scaling in the frame of the {\it local} magnetic field 
\citep{LV99,CV00}
and the nonlinear dynamics of MHD turbulence.

The theoretical understanding of MHD turbulence based on the GS95 model significantly advances the 
study of astroparticle physics. 
By using the scalings of compressible MHD turbulence numerically derived by 
\citet{CLV_incomp,CL03}, 
\citet{YL02,YL04} 
demonstrated that Alfv\'{e}n modes are inefficient in scattering cosmic rays (CRs) due to their scale-dependent anisotropy
(see also \citealt{Chan00}), 
while fast modes dominate the CR scattering in the diffuse ISM. 
Besides, 
pioneering works on the acceleration of dust grains in MHD turbulence have been carried out by 
\citet{LY02,YL03,YLD04,HLS12}, 
where MHD turbulence was identified as an important agent of grain acceleration.

The turbulence nature of magnetic fields entails the broadening of quasi-linear resonances. 
Attempts to modify the QLT to incorporate nonlinear effects are numerous
(e.g., \citealt{Volk:1973,Jones:1973,Goldstein:1976,Sha04}.
\citet{YL08} (hereafter YL08)
first performed the nonlinear extension of QLT in the framework of a realistic and numerically tested description of MHD turbulence. 
They found that the 
broadening effect due to the perturbations of particle orbits is essential for calculating the scattering and mean free paths of CRs.

In addition, the role of nonlinear dynamics of MHD turbulence in resonance broadening has been investigated by 
\citet{Chan00,Ly12,Lyn14}. 
By considering the broadening effect due to the decorrelation of turbulent magnetic fields
\citep{Ly12},
\citet{Lyn14}
argued the dominance of slow modes of MHD turbulence in particle acceleration.

To identify the physical regimes for the dominance of different broadening effects, 
a generalized broadened resonance function should be constructed incorporating both 
the geometric and dynamical properties of MHD turbulence. 
In this work, we adopt the up-to-date model of MHD turbulence and 
focus on the pitch-angle scattering of particles including both the transit time damping (TTD) and gyroresonance interactions
\citep{Schlickeiser02}.
To fully capture the physics of resonance broadening and have a general applicability of the study, 
we formulate the pitch-angle diffusion coefficients with a 
broadened resonance function applicable to a wide range of particle energies, turbulence conditions, and plasma parameters. 

In this paper, Section 2 contains first the analysis on the resonance broadening
and then calculations of the pitch-angle diffusion coefficients for TTD with slow and fast modes, respectively. 
Their relative importance in different physical conditions is also examined. 
In Section 3, we present the resonance-broadened gyroresonance with Alfv\'{e}n, slow, and fast modes.  
Final conclusions follow in Section 4.

\section{Resonance-broadened TTD}

As the magnetic analogue of Landau damping
\citep{Sum00},
TTD is the resonant interaction of particles with compressive waves.  
The resulting scattering process can be described with the pitch-angle diffusion coefficient 
\citep{Kulsrud_Pearce,Volk:1975},
\begin{equation}\label{eq:gel}
     D_{\mu\mu,T} = C_\mu \int d^3k \frac{k_\|^2}{k^2} [J_0^\prime (x)]^2 I(k) R(k)  , 
\end{equation}
where 
\begin{equation}\label{eq: cmu}
    C_\mu  = (1 - \mu^2) \frac{\Omega^2}{B_0^2}, 
\end{equation}
\begin{equation} \label{eq: asux}
     x = \frac{k_\perp v_\perp}{ \Omega}  = \frac{k_\perp }{r_g^{-1} }, 
\end{equation}
with the gyrofrequency $\Omega= |q| B_0 / \gamma mc$ for relativistic particles and 
$\Omega = |q| B_0 / mc$ for non-relativistic particles, 
the mean magnetic field $B_0$,
the gyroradius $r_g = v_\perp / \Omega$, 
the electric charge $q$, the mass $m$, and the Lorentz factor $\gamma$ of the particle, 
the light speed $c$, the pitch-angle cosine $\mu = v_\| / v$, 
particle speed $v$ and its parallel and perpendicular components $v_\|$, $v_\perp$ with respect to the magnetic field, 
wavenumber $k$ and its parallel and perpendicular components $k_\|$ and $k_\perp$. 
Besides, 
$J_0^\prime (x)$ is the derivative of the Bessel function, 
$I(k)$ is the power spectrum of magnetic field fluctuations, 
and $R(k)$ is the resonance function.

\subsection{Resonance broadening}
\label{ssec: rebr}

The standard QLT was formulated under the condition of sufficiently small turbulence amplitude
\citep{Jokipii1966,Pin52}. 
In such weak plasma turbulence, the assumption of unperturbed particle orbits is valid.

Regarding the TTD resonance,  
QLT requires that $v_\|$ should be matched to the wave parallel phase speed $\omega/k_\|$ for the particle to resonate with the wave,
where $\omega$ is the wave frequency.
This resonance condition corresponds to a linear resonance function
\begin{equation}\label{eq: ltb}
    R_L = \pi \delta(\omega_k - v_\| k_\| )  , 
\end{equation}
resulting in a singular diffusion process and inefficient particle scattering. 
It entails a threshold particle speed for resonance, 
\begin{equation}\label{eq: thr}
     v \geq V_\text{ph}, 
\end{equation}
where $V_\text{ph}$ is the phase speed of the wave. 
It implies that the particles with the speed below the threshold speed cannot undergo the TTD resonance.

However, the restricted condition of the QLT cannot be met in most astrophysical environments, where the strong MHD turbulence is present. 
We note that the concept of strong turbulence discussed here is different from the strong plasma turbulence with 
the particle kinetic energy comparable to the wave energy.
The strong MHD turbulence results from the nonlinear interaction of Alfv\'{e}n waves.
It is characterized by the critical balance between 
the turbulent eddy-turnover time in the direction perpendicular to the {\it local} magnetic field 
and the Alfv\'{e}n wave-crossing time in the direction parallel to the {\it local} magnetic field 
on every length scale within the inertial range of turbulence
\citep{GS95, LV99}.
As the consequences of the critical balance, 
(i) Alfv\'{e}n turbulence has a scale-dependent anisotropy, with more elongated turbulent eddies toward smaller scales
\citep{CV00,MG01,CLV_incomp,CL03}, and 
(ii) the turbulent eddy (or the wave packet) has its lifetime given by the eddy-turnover time (or the Alfv\'{e}n wave-crossing time).

Pseudo-Alfv\'{e}n modes in incompressible MHD turbulence and 
slow modes in compressible MHD turbulence have their dynamics governed by Alfv\'{e}n modes and 
have the same anisotropic scaling as Alfv\'{e}n modes
\citep{LG01,CL05}.
By contrast, fast modes in compressible MHD turbulence have isotropic scaling and independent energy cascade with a much slower 
cascading rate compared to other modes
\citep{CL02_PRL}.

Due to the parallel magnetic perturbations induced by pseudo-Alfv\'{e}n modes or compressible fast and slow modes, 
particles moving along the fluctuating magnetic field have a variation in $v_\|$  
\citep{Volk:1975},
\begin{equation}\label{eq: volk}
     \Delta v_\|  \approx v_\perp \Bigg(\frac{\langle\delta B_\|^2\rangle}{B_0^2}\Bigg)^\frac{1}{4} ,
\end{equation}
where $\delta B_\|$ represents the parallel magnetic perturbation. 
It arises due to the conservation of the 
magnetic moment $J_1 \propto v_\perp^2 / B$
when the magnetic field variation over a gyro period is insignificant. 
The existence of $ \Delta v_\|$ gives rise to the deviation from the unperturbed orbit and thus the broadening of 
the wave-particle resonance 
(YL08).

Another source of broadening is the dynamical decorrelation of MHD turbulence 
\citep{Ly12}.
It leads to the deviation from the linearized particle trajectory in the direction perpendicular to the {\it local} magnetic field 
with the characteristic timescale as the turbulent cascading time $\omega_\text{tur}^{-1}$.

By taking into account both broadening effects as discussed above, 
we have a general form of the broadened resonance function
\begin{equation}\label{eq: genrb}
\begin{aligned}
    R_B & = \text{Re} \int_0^\infty dt \exp \Big[ i(\omega_k - v_\| k_\|) t -\frac{1}{2} ({\bf{k}}\cdot \delta {\bf{r}}(t))^2\Big]  \\
            & = \text{Re} \int_0^\infty dt \exp \Big[ i(\omega_k - v_\| k_\|) t -\frac{1}{2} (k_\| \Delta v_\| + \omega_\text{tur})^2 t^2\Big] \\
            & = \frac{\sqrt{2\pi}}{2 (\Delta v_\| k_\|  + \omega_\text{tur})}
    \exp{\Bigg[-\frac{(\omega_k - v_\| k_\|)^2 }{2 ( \Delta v_\| k_\|+ \omega_\text{tur})^2}\Bigg]} .
\end{aligned}
\end{equation}
The two terms $\Delta v_\| k_\|$ and $\omega_\text{tur}$ come from the broadening effects due to 
magnetic perturbations and the dynamical decorrelation of MHD turbulence, respectively. 
We can easily see that high-energy particles with $v\gg V_\text{ph}$ mainly sample magnetic fluctuations when propagating through 
the quasistatic magnetic field. 
By contrast, 
low-energy particles with $v \ll V_\text{ph}$
can sample uncorrelated magnetic fields when being quasistatic relative to turbulent magnetic fields. 
More specifically, when 
\begin{equation}\label{eq: c1}
      \Delta v_\| k_\| \gg \omega_\text{tur} ,
\end{equation}
the broadening is primarily caused by the propagation of particles and magnetic perturbations. 
Accordingly, the general expression of $R_B$ reduces to,
\begin{equation}\label{eq: vbmf}
    R_{B1} = \frac{\sqrt{2\pi}}{2  \Delta v_\| k_\|}
    \exp{\Bigg[-\frac{(\omega_k - v_\| k_\|)^2 }{2 ( \Delta v_\| k_\|)^2}\Bigg]} .
\end{equation}
A similar form of resonance function was introduced in YL08 for studying the scattering of CRs. 
In the opposite limit of 
\begin{equation}\label{eq: c2}
      \Delta v_\| k_\| \ll \omega_\text{tur},
\end{equation}
the broadening is dominated by the decorrelation of turbulent magnetic fields, and $R_B$ approximately becomes 
\begin{equation}\label{eq: rb2}
    R_{B2} = \frac{\sqrt{2\pi}}{2 \omega_\text{tur}}
    \exp{\Bigg[-\frac{(\omega_k - v_\| k_\|)^2 }{2 \omega_\text{tur}^2}\Bigg]} .
\end{equation}
A similar Gaussian resonance function involving wave decoherence was earlier proposed by
\citet{Ly12}.
We will next clarify the relative importance between different broadening effects and the applicability of $R_{B1}$ 
and $R_{B2}$ in different regimes.

\subsection{Pitch-angle diffusion coefficients}

Compressible MHD turbulence can be decomposed into Alfv\'{e}n, slow, and fast modes 
\citep{CL03}.
Both compressible slow and fast modes contribute to the TTD scattering. 
The corresponding $D_{\mu\mu,T}$ has two components associated with slow and fast modes, respectively,  
\begin{equation}\label{eq: totdut}
     D_{\mu\mu,T} = D_{\mu\mu,s,T} + D_{\mu\mu,f,T}. 
\end{equation}
In what follows, we separately calculate $D_{\mu\mu,s,T} $ and $D_{\mu\mu,f,T} $ with resonance broadening, 
and then compare their relative importance under different conditions.

\subsubsection{Slow modes}

As mentioned above, 
slow modes have the same anisotropic scaling as Alfv\'{e}n modes, 
which is reflected in their magnetic energy spectrum  
\citep{CLV_incomp},
\begin{equation}\label{eq: slspe}
     I_s(k) = C_s  k_\perp^{-\frac{10}{3}} \exp{\Bigg(-L^\frac{1}{3}\frac{k_\|}{k_\perp^\frac{2}{3}}\Bigg)},
\end{equation} 
where $L$ is the injection scale of strong MHD turbulence. 
The normalization factor is 
\begin{equation}\label{eq: cs}
    C_s = \frac{1}{6 \pi} \delta B_s^2 L^{-\frac{1}{3}}, 
\end{equation}
and thus 
\begin{equation}
     \frac{\delta B_s^2}{2} = \int d^3 k I_s(k),
\end{equation}
where $\delta B_s$ is the rms strength of the fluctuating magnetic fields induced by slow modes. 
$\delta B_s$ depends on the ratio of gas pressure to magnetic pressure 
$\beta = 2 c_s^2 / V_A^2$, where $c_s$ is the sound speed, and $V_A = B_0 / \sqrt{4\pi \rho}$ is the Alfv\'{e}n speed with the density $\rho$. 
Approximately, there is 
\citep{CL03}
\begin{equation}\label{eq: cmsh}
     \frac{\delta B_s}{B_0} \sim \frac{V_{Ls}}{V_A},
\end{equation}
in the high-$\beta$ case, and 
\begin{equation}\label{eq: cmsl}
      \frac{\delta B_s}{B_0} \sim \frac{c_s V_{Ls}}{V_A^2}.
\end{equation}
in the low-$\beta$ case. Here $V_{Ls}$ is the turbulent speed of slow modes at $L$.
It shows that the magnetic fluctuations of slow modes in a high-$\beta$ medium are much larger than those in a low-$\beta$ medium.

On the other hand, the wave-like motions of slow modes propagate along the magnetic field, so there is 
\begin{equation}
    \omega_k = V_\text{ph} k_\|, 
\end{equation}
where $V_\text{ph} \approx V_A$ in the high-$\beta$ case, and 
$V_\text{ph} \approx c_s$ in the low-$\beta$ case.
In what follows we analytically calculate the pitch-angle diffusion coefficients of slow modes 
with the linear and broadened resonance functions.

(1) Linear resonance  

By inserting Eqs. \eqref{eq: ltb} and \eqref{eq: slspe} into Eq. \eqref{eq:gel}, 
we can derive the pitch-angle diffusion coefficient of slow modes by using the discrete resonance of the QLT,  
\begin{equation}\label{eq: duusl}
\begin{aligned}
   D_{\mu\mu,sL,T} & =  \frac{\pi^2}{2} C_\mu C_s   \frac{v_\perp^2}{\Omega^2} \delta(V_\text{ph}-v_\|)  \\
   & \int_{L^{-1}}^{k_{\perp,\text{max}}} k_\perp^{-\frac{7}{3}} \int_{L^{-1}}^{k_{\|,\text{max}}} k_\| \exp (- L^\frac{1}{3} k_\perp^{-\frac{2}{3}} k_\|) dk_\| dk_\perp    \\
   &  = \frac{\pi^2}{2} \frac{C_s}{B_0^2} L^{-\frac{2}{3}}  \ln \Big(\frac{L}{l_{\perp,\text{min}}}\Big) v^2 
     (1-\mu^2)^2  \delta(v_\| - V_\text{ph}).
\end{aligned}
\end{equation}
Since particles are insensitive to the averaged magnetic fields on length scales below $r_g$, 
we only consider the magnetic fluctuations on scales larger than $r_g$ and thus $x<1$ as a simplifying assumption. 
At $x<1$, there is 
\begin{equation}\label{eq: simbel}
    [J_0^\prime(x)]^2 = [J_1(x)]^2 \sim \frac{x^2}{4}.
\end{equation}
Correspondingly, the upper limit of integration in Eq. \eqref{eq:gel} is determined by the larger value between the dissipation scale $l_d$ of magnetic fluctuations 
and $r_g$, i.e., $l_{\perp,\text{min}} = 1/ k_{\perp,\text{max}} = \text{max}[l_d, r_g]$. 
In addition, since TTD is dominated by the interaction of particles with 
the magnetic fluctuations at $l_{\perp,\text{min}}$, 
where the turbulence anisotropy is prominent with $k_\perp \gg k_\|$, 
we assume $k^2 = k_\perp^2 + k_\|^2 \sim k_\perp^2$ when calculating the above integral.

The above result clearly shows that only the particles exactly matching the linear resonance condition can be scattered.

(2) Broadened resonance 

With the use of $R_B$ in Eq. \eqref{eq: genrb},
the resonance-broadened diffusion coefficient can be expressed,
\begin{equation}\label{eq: gendus}
\begin{aligned}
 D_{\mu\mu,s,T} =& \frac{\sqrt{2\pi}}{8}C_\mu C_s  \frac{v_\perp^2}{\Omega^2} (\Delta v_\|+V_A)^{-1}  \exp \Bigg[ - \frac{(V_\text{ph}-v_\|)^2}{2 (\Delta v_\| + V_A)^2 }\Bigg] \\
&  \int d^3k  \frac{k_\|}{k_\|^2 + k_\perp^2} k_\perp^{-\frac{4}{3}} \exp \Bigg[-L^\frac{1}{3} \frac{k_\|}{k_\perp^\frac{2}{3}}  \Bigg]   .
\end{aligned} 
\end{equation}
Here we adopt 
\begin{equation}
    \omega_\text{tur,s} =  V_A k_\|
\end{equation}
based on the critical balance of strong MHD turbulence. 
In Appendix \ref{aa1: tur}, we also provide the analysis with $\omega_\text{tur,s}$ given by the eddy-turnover rate. 
We further derive its approximate analytical result by using the same simplifying assumptions 
for deriving Eq. \eqref{eq: duusl},
\begin{equation}\label{eq: sange}
\begin{aligned}
    D_{\mu\mu,s,T} \approx &\frac{\sqrt{2}}{4} \pi^\frac{3}{2} \frac{C_s}{B_0^2} (\Delta v_\|+V_A)^{-1} L^{-\frac{2}{3}} \ln \Big(\frac{L}{l_{\perp,\text{min}}}\Big) v^2   \\
   & (1-\mu^2)^2   \exp \Bigg[ - \frac{(V_\text{ph}-v_\|)^2}{2 (\Delta v_\| + V_A)^2 }\Bigg] .
\end{aligned}
\end{equation}
By comparing with $D_{\mu\mu,sL,T}$ in Eq. \eqref{eq: duusl}, 
we see that the delta function of $v_\|$ in $D_{\mu\mu,sL,T}$ is replaced by a Gaussian function with the 
variance $(\Delta v_\| + V_A)^2$.
With the broadened resonance, 
particles with a broad range of $v$ including $v< V_\text{ph}$ can be effectively scattered.

$\Delta v_\| $ and $V_A$ in $D_{\mu\mu,s,T}$ correspond to the two broadening effects as discussed in Section \ref{ssec: rebr}.
For high-energy particles, $\Delta v_\| (\propto v_\perp$, Eq. \eqref{eq: volk}) tends to be much larger than $V_A$.
The broadening due to magnetic perturbations is dominant, and the general form of $D_{\mu\mu,s,T}$ in Eq. \eqref{eq: sange} can be 
simplified as 
\begin{equation}\label{eq: ssbf}
\begin{aligned}
     D_{\mu\mu,s1,T} \approx & \frac{\sqrt{2}}{4} \pi^\frac{3}{2} \frac{C_s}{B_0^2} \Bigg(\frac{\langle\delta B_\|^2\rangle}{B_0^2}\Bigg)^{-\frac{1}{4}}
     L^{-\frac{2}{3}} \ln \Big(\frac{L}{l_{\perp,\text{min}}}\Big) v  \\
  &   (1-\mu^2)^\frac{3}{2} \exp \Bigg[-\frac{(V_\text{ph}-v_\|)^2}{2\Delta v_\|^2}\Bigg] ,
\end{aligned}
\end{equation}
where Eq. \eqref{eq: volk} is inserted.
Since slow modes have $V_\text{ph} \leq V_A$, for high-energy particles the above equation can be further simplified as 
\begin{equation}\label{eq: sbfurs}
\begin{aligned}
     D_{\mu\mu,s1,T} \approx & \frac{\sqrt{2}}{4} \pi^\frac{3}{2} \frac{C_s}{B_0^2} \Bigg(\frac{\langle\delta B_\|^2\rangle}{B_0^2}\Bigg)^{-\frac{1}{4}}
     L^{-\frac{2}{3}} \ln \Big(\frac{L}{l_{\perp,\text{min}}}\Big) v  \\
  &   (1-\mu^2)^\frac{3}{2} \exp \Bigg[-\frac{v_\|^2}{2\Delta v_\|^2}\Bigg] ,
\end{aligned}
\end{equation}
which is valid for most pitch angles except for $\mu \rightarrow 0$. 
With a considerable level of magnetic perturbations,
$\Delta v_\|$ and $v_\|$ are of the same order of magnitude, 
so the exponential term in the above equation is not small. 
Besides, $D_{\mu\mu,s1,T}$ has a dependence on $\mu$ as 
\begin{equation}
    D_{\mu\mu,s1,T} \propto (1-\mu^2)^\frac{3}{2} \exp \Bigg[-\frac{ \mu^2}{2 \Big(\frac{\langle\delta B_\|^2\rangle}{B_0^2}\Big)^\frac{1}{2}(1-\mu^2)}  \Bigg] .
\end{equation}
So $D_{\mu\mu,s1,T}$ is approximately constant at a small $\mu$ and significantly drops 
when $\mu$ approaches $1$.

For sufficiently low-energy particles with $\Delta v_\| \ll V_A$, the broadening due to turbulent decorrelation is dominant. 
In this case $D_{\mu\mu,s,T}$ can be approximated by 
\begin{equation}\label{eq: dscb}
\begin{aligned}
    D_{\mu\mu,s2,T} \approx &\frac{\sqrt{2}}{4} \pi^\frac{3}{2} \frac{C_s}{B_0^2} V_A^{-1} L^{-\frac{2}{3}} \ln \Big(\frac{L}{l_{\perp,\text{min}}}\Big) v^2   \\
   & (1-\mu^2)^2  \exp \Bigg[ - \frac{(V_\text{ph}-v_\|)^2}{2 V_A^2 }\Bigg] ,
\end{aligned}
\end{equation}
which can be further cast as 
\begin{equation}\label{eq: dsfurs}
\begin{aligned}
    D_{\mu\mu,s2,T} \approx &\frac{\sqrt{2}}{4} \pi^\frac{3}{2} \frac{C_s}{B_0^2} V_A^{-1} L^{-\frac{2}{3}} \ln \Big(\frac{L}{l_{\perp,\text{min}}}\Big) v^2   \\
   & (1-\mu^2)^2  \exp \Bigg[ - \frac{V_\text{ph}^2}{2 V_A^2 }\Bigg] 
\end{aligned}
\end{equation}
when $v_\| \ll V_\text{ph}$.
With $V_\text{ph} \leq V_A$ for slow modes, the exponential term is of the order of unity. 

The above analysis shows that 
owing to the broadened resonance, both high- and low-energy particles can be effectively scattered at pitch angles away from $\mu = 1$.

\subsubsection{Fast modes}\label{ssec: fatd}

Fast modes are weakly coupled with Alfv\'{e}n and slow modes.
They have an isotropic scaling and an energy spectrum 
\citep{CL02_PRL}
\begin{equation}\label{eq: fsep}
     I_f(k) = C_f k^{-\frac{7}{2}}.
\end{equation}
The normalization constant is 
\begin{equation}\label{eq: cf}
    C_f = \frac{1}{16 \pi} \delta B_f^2 L^{-\frac{1}{2}}, 
\end{equation}
so that there is 
\begin{equation}
     \frac{\delta B_f^2}{2} = \int d^3  k I_f(k).
\end{equation}
The rms strength of magnetic fluctuations of fast modes also depends on the plasma $\beta$. 
It is approximately given by 
\citep{CL03}
\begin{equation}
    \frac{\delta B_f}{B_0} \sim \frac{V_{Lf}}{c_s},
\end{equation}
in a high-$\beta$ medium, and 
\begin{equation}
    \frac{\delta B_f}{B_0} \sim \frac{V_{Lf}}{V_A}.
\end{equation}
in a low-$\beta$ medium, where $V_{Lf}$ is the turbulent speed of fast modes at $L$. 
By comparing with $\delta B_s$ in Eqs. \eqref{eq: cmsh} and \eqref{eq: cmsl}, 
we find that provided $V_{Ls} \sim V_{Lf}$, 
$\delta B_f$ is much smaller than $\delta B_s$ in a high-$\beta$ medium,
but much larger than $\delta B_s$ in a low-$\beta$ medium.

Besides, fast modes have the dispersion relation as
\begin{equation}
  \omega_k = V_\text{ph}k, 
\end{equation}
where $V_\text{ph} \approx c_s$ in a high-$\beta$ medium, and 
$V_\text{ph} \approx V_A$ in a low-$\beta$ medium. 
Next, we will investigate the pitch-angle diffusion coefficients of fast modes and the effect of 
resonance broadening on particle scattering.

(1) Linear resonance

The diffusion coefficient with the linear resonance function is (Eqs. \eqref{eq:gel}, \eqref{eq: ltb}, and \eqref{eq: fsep})
\begin{equation}
\begin{aligned}
    D_{\mu\mu,fL,T} &= \pi^2 C_\mu C_f \frac{v_\perp^2}{\Omega^2} \frac{1}{v_\|} \\
   & \int_{L^{-1}}^{k_\text{max}} k^{-\frac{1}{2}} \int_0^1  \eta^2 (1-\eta^2) \delta\bigg(\eta - \frac{V_\text{ph}}{v_\|}\bigg) d \eta dk  ,
\end{aligned}
\end{equation}
where $\eta = \cos\theta$, and $\theta$ is the angle between $\bf{k}$ and $\bf{B}$. 
We again use the approximation in Eq. \eqref{eq: simbel} and obtain the analytical expression 
\begin{equation}\label{eq: ftlb}
\begin{aligned}
 D_{\mu\mu,fL,T} 
                         = & 2H^* \pi^2 \frac{C_f}{B_0^2} \Big(l_\text{min}^{-\frac{1}{2}} - L^{-\frac{1}{2}}\Big) v \\
                            & \frac{(1-\mu^2)^2}{\mu}  \bigg(\frac{V_\text{ph}}{v_\|}\bigg)^2 \bigg[1-\bigg(\frac{V_\text{ph}}{v_\|}\bigg)^2\bigg]  ,
\end{aligned}
\end{equation}
where $l_\text{min} = 1/k_\text{max} = \text{max}[l_d, r_g]$, and 
\begin{subnumcases}
 {H^* = H\bigg(1-\frac{V_\text{ph}}{v_\|}\bigg) H\bigg(\frac{V_\text{ph}}{v_\|}\bigg)=}
         0,  ~v_\| < V_\text{ph},  \\
         \frac{1}{2},  v_\|=V_\text{ph},  \\
         1,  ~v_\|>V_\text{ph},
\end{subnumcases}
with the Heaviside step function $H$.
Different from $D_{\mu\mu,sL,T}$ (Eq. \eqref{eq: duusl}), $D_{\mu\mu,fL,T}$ does not contain a delta function. 
The particles satisfying $v_\| \geq V_\text{ph}$ can all be scattered. 
When $v_\| \gg V_\text{ph}$, it scales with $\mu$ as 
\begin{equation}
      D_{\mu\mu,fL,T} \propto \frac{(1-\mu^2)^2}{\mu^3}. 
\end{equation}
$D_{\mu\mu,fL,T}$ decreases with $\mu$ as $\mu^{-3}$ at a relatively small $\mu$.
Moreover, different from $D_{\mu\mu,sL,T}\propto \ln (L/l_{\perp,\text{min}})$,
$D_{\mu\mu,fL,T}$ depends on $(L/l_\text{min})^{1/2}$.

(2) Broadened resonance

The turbulent cascade of fast modes is slow, with the cascading rate much smaller than that of Alfv\'{e}n and slow modes 
\citep{CL02_PRL},
\begin{equation}\label{eq: casrfa}
   \omega_\text{tur,f} = \frac{v_k}{V_\text{ph}} (k v_k) = 
   \frac{V_{Lf}^2}{V_\text{ph}} L^{-\frac{1}{2}} k^\frac{1}{2}.
\end{equation}
We first consider the case of high-energy particles. 
As there generally exists $\omega_\text{tur,s} > \omega_\text{tur,f}$, if there is 
\begin{equation}
      k_\| \Delta v_\| \gg \omega_\text{tur,s}
\end{equation}
for slow modes, then the condition for the dominance of the broadening due to magnetic perturbations is naturally satisfied for fast modes.
The corresponding diffusion coefficient is (Eqs. \eqref{eq:gel}, \eqref{eq: vbmf}, and \eqref{eq: fsep}), 
\begin{equation}\label{eq: rodfmf}
\begin{aligned}
 D_{\mu\mu,f1,T} \approx &\frac{\sqrt{2}}{4}\pi^\frac{3}{2} C_\mu C_f \frac{v_\perp^2}{\Omega^2} \frac{1}{\Delta v_\|} \\
& \int k^{-\frac{1}{2}} \cos\theta \sin^3\theta
 \exp \Bigg[- \frac{(kV_\text{ph} - k_\| v_\|)^2}{2k_\|^2 \Delta v_\|^2}\Bigg]  .
\end{aligned}
\end{equation}
For high-energy particles with $v_\| \gg V_\text{ph}$, from the above equation we can approximately derive 
\begin{equation}\label{eq: fbma}
\begin{aligned}
 D_{\mu\mu,f1,T} \approx &\frac{\sqrt{2}}{2}\pi^\frac{3}{2} C_\mu C_f \frac{v_\perp^2}{\Omega^2} \frac{1}{\Delta v_\|} 
 \exp \Bigg[- \frac{  v_\|^2}{2\Delta v_\|^2}\Bigg]    \\
& \int_{L^{-1}}^{k_\text{max}} k^{-\frac{1}{2}} \int_0^1 \zeta^3  d\zeta dk \\
= & \frac{\sqrt{2}}{4}\pi^\frac{3}{2}  \frac{C_f}{B_0^2}  \Bigg(\frac{\langle\delta B_\|^2\rangle}{B_0^2}\Bigg)^{-\frac{1}{4}} \Big(l_\text{min}^{-\frac{1}{2}} - L^{-\frac{1}{2}}\Big)  v  \\
   & (1-\mu^2)^\frac{3}{2} \exp \Bigg[- \frac{  v_\|^2}{2\Delta v_\|^2}\Bigg]  ,
\end{aligned}
\end{equation}
where $\zeta = \sin\theta$.
By comparing with $D_{\mu\mu,fL,T}$ in Eq. \eqref{eq: ftlb}, which is proportional to $(V_\text{ph}/v_\|)^2$ and thus 
has a small value when $v_\| \gg V_\text{ph}$, 
$D_{\mu\mu,f1,T}$ depends on $\exp (-   v_\|^2/2\Delta v_\|^2)$ because of the broadening effect, and consequently
high-energy particles can be much more effectively scattered.

Compared to $D_{\mu\mu,s1,T}$ in Eq. \eqref{eq: sbfurs}, we see that similar to the case with the linear resonance, 
$D_{\mu\mu,f1,T}$ depends on $(L/l_\text{min})^{1/2}$, while 
$D_{\mu\mu,s1,T}$ depends on $\ln (L/l_{\perp,\text{min}})$. 
This difference is not related to the broadening effect. 
The logarithmic dependence on $l_{\perp,\text{min}}$ of 
$D_{\mu\mu,sL,T}$ and $D_{\mu\mu,s1,T}$ comes from the anisotropy of slow modes. 
The energy distribution is increasingly anisotropic toward small scales. With insignificant magnetic fluctuations parallel to the magnetic field on 
small scales, TTD with slow modes tends to be less effective compared to that with fast modes. 
As a test, in Appendix \ref{asec: fs}, we adopt an isotropic spectrum but with the same spectral index as that of slow modes 
to calculate the diffusion coefficient of fast modes. 
It again shows a power-law dependence on $l_\text{min}$, but with a different scaling due to the different spectral index used.

For sufficiently low-energy particles, there can be (Eq. \eqref{eq: c2}),
\begin{equation}\label{eq: dislc}
      k_\| \Delta v_\| \ll \omega_\text{tur,f}.
\end{equation}
Then the broadening is dominated by the turbulent decorrelation, and 
the diffusion coefficient is (Eqs. \eqref{eq:gel}, \eqref{eq: rb2}, and \eqref{eq: fsep})
\begin{equation}
\begin{aligned}
    D_{\mu\mu,f2,T} \approx &\frac{\sqrt{2}}{4} \pi^\frac{3}{2} C_\mu C_f \frac{V_\text{ph}}{V_{Lf}^2} L^\frac{1}{2} \frac{v_\perp^2}{\Omega^2} \\
   & \int \cos^2\theta \sin^3\theta \exp{\bigg[-\frac{(kV_\text{ph} - k_\| v_\|)^2}{2 V_{Lf}^4 V_\text{ph}^{-2} L^{-1} k}\bigg]}  .
\end{aligned}
\end{equation}
We consider $v_\| \ll V_\text{ph}$ and approximately have 
\begin{equation}\label{eq: leftdd}
\begin{aligned}
    D_{\mu\mu,f2,T} \approx &\frac{\sqrt{2}}{2} \pi^\frac{3}{2} C_\mu C_f \frac{V_\text{ph}}{V_{Lf}^2} L^\frac{1}{2} \frac{v_\perp^2}{\Omega^2} \\
   & \int_{L^{-1}}^{k_\text{max}} \int_0^1 \eta^2 (1- \eta^2) d \eta \exp{\bigg[-\frac{V_\text{ph}^4 }{2 V_{Lf}^4  L^{-1} } k\bigg]} dk \\
   \approx & \frac{2\sqrt{2}}{15} \pi^\frac{3}{2} \frac{C_f}{B_0^2} \frac{V_{Lf}^2}{V_\text{ph}^3} L^{-\frac{1}{2}} v^2
     (1-\mu^2)^2 \exp{\bigg[-\frac{V_\text{ph}^4 }{2 V_{Lf}^4  }\bigg]} .
\end{aligned}
\end{equation}
As the main contribution to the integral comes from $k=1/L$, the consequent $D_{\mu\mu,f2,T}$ does not depend on $l_\text{min}$.

Due to the particularly slow cascade of fast modes and the scaling $\omega_\text{tur,f} \propto k^{1/2}$,
it is likely that the condition in Eq. \eqref{eq: dislc} breaks down on large wavenumbers. 
In this case we should use the general form of the broadened resonance function $R_B$ in Eq. \eqref{eq: genrb}.
At $v_\| \ll V_\text{ph}$, $R_B$ is 
\begin{equation}
    R_B = \frac{\sqrt{2\pi}}{2 (k_\| \Delta v_\| + \omega_\text{tur,f})}
    \exp{\Bigg[-\frac{V_\text{ph}^2 k ^2 }{2 (k_\| \Delta v_\| + \omega_\text{tur,f})^2}\Bigg]} ,
\end{equation}
which is approximately 
\begin{equation}
    R_B \approx \frac{\sqrt{2\pi}}{2  \omega_\text{tur,f}}
    \exp{\Bigg[-\frac{V_\text{ph}^2 k ^2 }{2  \omega_\text{tur,f}^2}\Bigg]} 
\end{equation}
at small wavenumbers, and 
\begin{equation}
    R_B \approx \frac{\sqrt{2\pi}}{2 k_\| \Delta v_\| }
    \exp{\Bigg[-\frac{V_\text{ph}^2 k ^2 }{2 (k_\| \Delta v_\| )^2}\Bigg]} 
\end{equation}
at large wavenumbers. 
Obviously, $R_B$ at large wavenumbers is small due to $\Delta v_\| \ll V_\text{ph}$ for very low-energy particles, 
while $R_B$ at small wavenumbers decreases with increasing $k$. 
So even when the condition in Eq. \eqref{eq: dislc} cannot hold on large wavenumbers, 
$k=1/L$ still mainly contributes to $D_{\mu\mu,f2,T}$, and its expression in 
Eq. \eqref{eq: leftdd} is still valid for sufficiently low-energy particles.

Owing to the broadening effect, particles in the range $v_\| < V_\text{ph}$ can also be scattered with a nonzero pitch-angle diffusion coefficient. 
In the limiting case with $v_\| \ll V_\text{ph}$, $D_{\mu\mu,f2,T}$ (Eq. \eqref{eq: leftdd}) does not have the dependence on $(L/l_\text{min})^{1/2}$ 
when the broadening is dominated by the turbulent decorrelation effect.

\begin{table*}[!htbp]
\renewcommand\arraystretch{1.5}
\centering
\begin{threeparttable}
\caption[]{Comparison between $D_{\mu\mu,T}$ with linear resonance and broadened resonance}\label{tab:vel} 
  \begin{tabular}{c|c|c|c|c}
     \toprule
\multirow{2}*{}       &       \multicolumn{2}{c|}{Slow modes}                      &  \multicolumn{2}{c}{Fast modes}   \\
                                               \cline{2-5}
                                            &             $v_\|\ll V_\text{ph}$     & $v_\| \gg V_\text{ph}$    &  $v_\| \ll V_\text{ph}$     & $v_\| \gg V_\text{ph}$   \\
                      \hline
  Linear resonance              &  $0$   &  $0$      &   $0$   & $\propto v^{-1}$ (Eq. \eqref{eq: ftlb}) \\
                      \hline
  Broadened resonance                 &  $\propto v^2$ (Eq. \eqref{eq: dsfurs})  &   $\propto v$ (Eq. \eqref{eq: sbfurs})    &  $\propto v^2$ (Eq. \eqref{eq: leftdd})   & $\propto v$ (Eq. \eqref{eq: fbma})\\
     \bottomrule
    \end{tabular}
 \end{threeparttable}
\end{table*}

In Table \ref{tab:vel}, we summarize $D_{\mu\mu,T}$ for both slow and fast modes, where 
the effect of resonance broadening on TTD scattering of both low- and high-energy particles can be clearly seen.

\subsubsection{Comparison between slow and fast modes}
\label{sssec: com}

With the above analytical expressions of the pitch-angle diffusion coefficients of the resonance-broadened TTD, 
we are able to examine the relative importance between slow and fast modes in TTD scattering of particles in different regimes.

(1) High-energy particles ($v_\| \gg V_\text{ph}$)

For high-energy particles satisfying the condition in Eq. \eqref{eq: c1}, 
the broadening due to magnetic fluctuations is dominant. 
The total $D_{\mu\mu,T}$ (Eq. \eqref{eq: totdut}) can be approximately given by (Eqs. \eqref{eq: sbfurs} and \eqref{eq: fbma})
\begin{equation}\label{eq: tothetd}
\begin{aligned}
   & D_{\mu\mu,T} \approx D_{\mu\mu,s1,T} + D_{\mu\mu,f1,T} \\
                           &= \frac{\sqrt{2}}{4} \pi^\frac{3}{2} \frac{1}{B_0^2} \Bigg(\frac{\langle\delta B_\|^2\rangle}{B_0^2}\Bigg)^{-\frac{1}{4}} v
                           (1-\mu^2)^\frac{3}{2} \exp \Bigg[-\frac{v_\|^2}{2\Delta v_\|^2}\Bigg] \\
          &~~~~~~  \Big[C_s L^{-\frac{2}{3}} \ln \Big(\frac{L}{l_{\perp,\text{min},s}}\Big) +  C_f   \Big(l_{\text{min},f}^{-\frac{1}{2}} - L^{-\frac{1}{2}}\Big) \Big] .
\end{aligned}
\end{equation}
To compare the relative importance between $D_{\mu\mu,s1,T}$ and $D_{\mu\mu,f1,T}$, from the above equation we find their ratio as 
(Eqs. \eqref{eq: cs} and \eqref{eq: cf})
\begin{equation}   
     \frac{D_{\mu\mu, s1,T}}{D_{\mu\mu, f1,T}}  = \frac{C_s L^{-\frac{2}{3}} \ln \big(\frac{L}{l_{\perp,\text{min},s}}\big)}{C_f\big(l_{\text{min},f}^{-\frac{1}{2}} - L^{-\frac{1}{2}}\big)}  
     =  \frac{8}{3} \frac{\delta B_s^2}{\delta B_f^2}
          \frac{   \ln \big(\frac{L}{l_{\perp,\text{min},s}}\big)}{ \sqrt{\frac{L}{l_{\text{min},f}}} - 1}   ,
\end{equation}
where we assume $v_\| \gg V_\text{ph}$. 
In terms of $\beta$, the above ratio is 
\begin{equation}\label{eq: genbefs}
     \frac{D_{\mu\mu, s1,T}}{D_{\mu\mu, f1,T}}  
     =  \frac{4}{3}  \frac{V_{Ls}^2}{V_{Lf}^2} 
          \frac{   \ln \big(\frac{L}{l_{\perp,\text{min},s}}\big)}{ \sqrt{\frac{L}{l_{\text{min},f}}} - 1}  \beta  
     \approx \frac{4}{3}  
          \frac{   \ln \big(\frac{L}{l_{\perp,\text{min},s}}\big)}{ \sqrt{\frac{L}{l_{\text{min},f}}} - 1}  \beta   
\end{equation}
in both high- and low-$\beta$ media, where we also assume that the turbulent energies contained in slow and fast modes are comparable, i.e., 
$V_{Ls} \approx V_{Lf}$.

\begin{figure*}[htbp]
\centering   

\subfigure[]{
   \includegraphics[width=8.5cm]{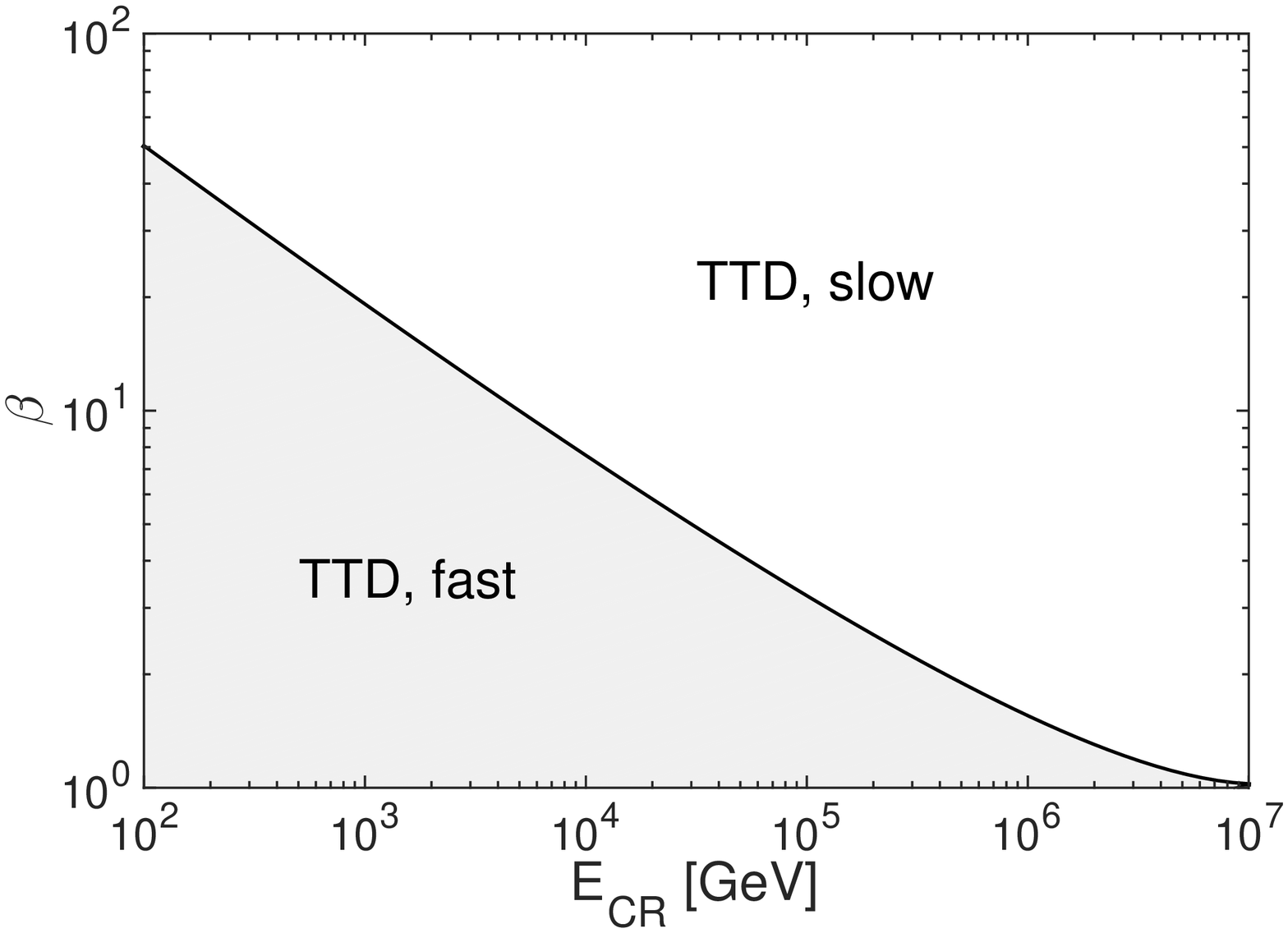}\label{fig: dube}}
\subfigure[]{
   \includegraphics[width=8.5cm]{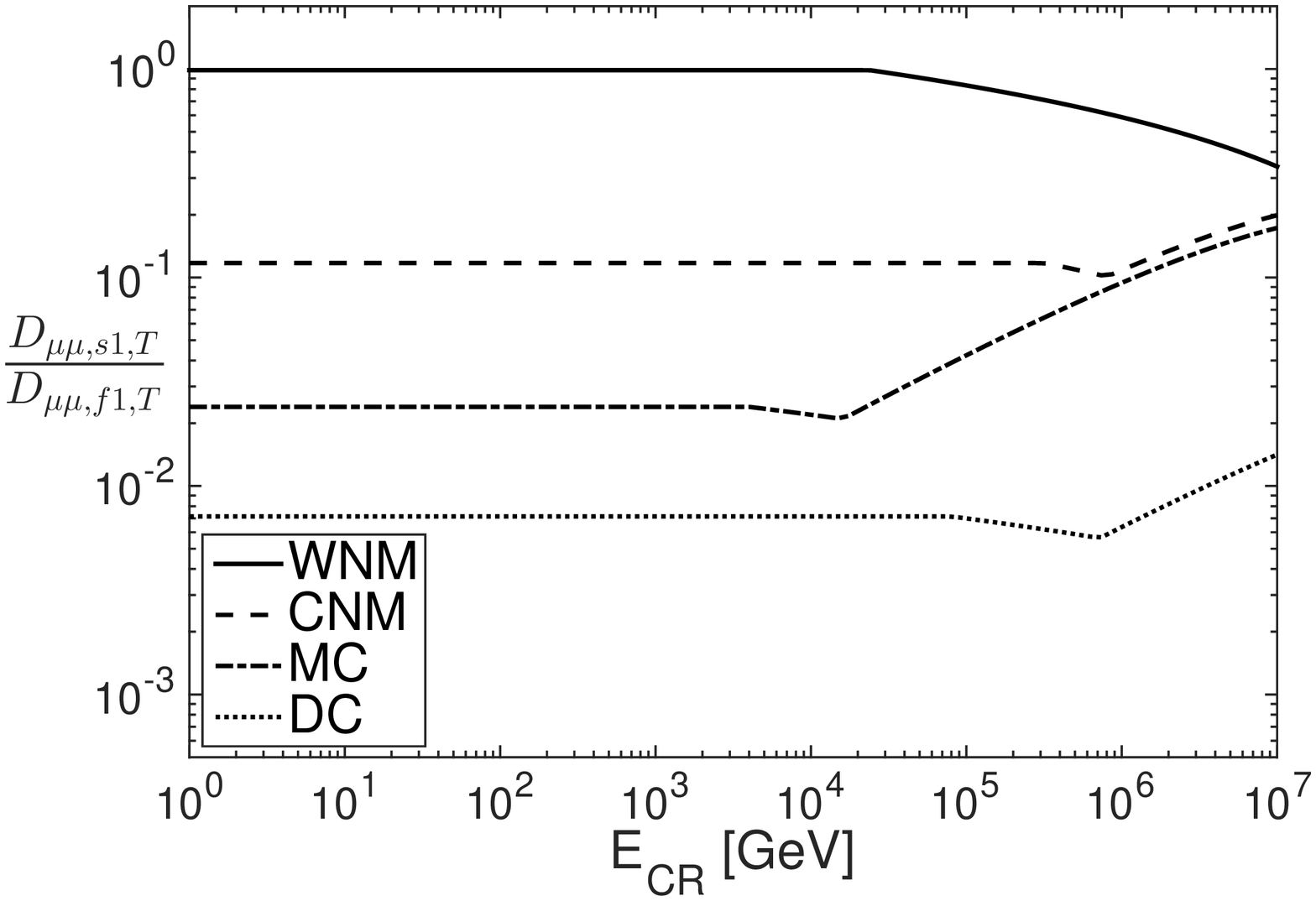}\label{fig: ducismfs}}

\caption{(a) The ranges of $\beta$ and $E_\text{CR}$ for the dominance of slow and fast modes in TTD resonance. 
The shaded region below the solid line corresponds to the relation in Eq. \eqref{eq: ttdbc}.
(b) $D_{\mu\mu, s1,T}/D_{\mu\mu, f1,T}$ vs. $E_\text{CR}$ for CR protons in partially ionized interstellar phases. }
\end{figure*}

For particles with $r_g$ larger than $l_d$ of both slow and fast modes, 
the ratio in Eq. \eqref{eq: genbefs} is in fact 
\begin{equation} \label{eq: hiecrat}
     \frac{D_{\mu\mu, s1,T}}{D_{\mu\mu, f1,T}}    
     \approx \frac{4}{3}  
          \frac{   \ln \big(\frac{L}{r_g}\big)}{ \sqrt{\frac{L}{r_g}} - 1}  \beta  
      \approx  \frac{4}{3}  
          \frac{   \ln \big(\frac{L}{r_g}\big)}{ \sqrt{\frac{L}{r_g}} }  \beta     
\end{equation}
under the condition $L \gg r_g$. 
We can then immediately see that when 
\begin{equation}\label{eq: ttdbc}
     \beta < \frac{3}{4} \frac{\sqrt{\frac{L}{r_g}} }{ \ln \big(\frac{L}{r_g}\big)}, 
\end{equation}
fast modes dominate the TTD resonance of energetic particles, 
while TTD with slow modes become more important at a larger $\beta$. 
The relation in Eq. \eqref{eq: ttdbc} is illustrated in Fig. \ref{fig: dube}, and 
$E_\text{CR}$ is the energy of CR protons corresponding to $r_g$ at $\mu = 0$. 
The parameters used are 
$L = 30$ pc and $B_0 = 3 ~\mu$ G.

In reality, the damping of turbulence should be taken into account. 
Fast modes are usually subjected to a strong damping effect and have a large $l_d$. 
For particles with $r_g$ larger than $l_{d,s}$ of slow modes but smaller than $l_{d,f}$ of fast modes, 
the ratio in Eq. \eqref{eq: genbefs} can be rewritten as 
\begin{equation}
     \frac{D_{\mu\mu, s1,T}}{D_{\mu\mu, f1,T}}    
     \approx \frac{4}{3}  
          \frac{   \ln \big(\frac{L}{r_g}\big)}{ \sqrt{\frac{L}{l_{d,f}}} }  \beta   
\end{equation}
when $L \gg l_{d,f}$.
If the particles have $r_g$ smaller than both $l_{d,s}$ and $l_{d,f}$, then there is 
\begin{equation}\label{eq: lowsrl}
     \frac{D_{\mu\mu, s1,T}}{D_{\mu\mu, f1,T}}    
     \approx \frac{4}{3}  
          \frac{   \ln \big(\frac{L}{l_{d,s}}\big)}{ \sqrt{\frac{L}{l_{d,f}}} }  \beta   
\end{equation}
when $L\gg l_{d,f}$.
Therefore, the above results can be summarized in a more general form,  
\begin{subnumcases}
{\frac{D_{\mu\mu, s1,T}}{D_{\mu\mu, f1,T}}\approx \label{eq: genradfs}} 
       \frac{4}{3} \frac{   \ln \big(\frac{L}{l_{d,s}}\big)}{ \sqrt{\frac{L}{l_{d,f}}} }  \beta , ~~~~~~~~~r_g < l_{d,s}, \\
       \frac{4}{3} \frac{   \ln \big(\frac{L}{r_g}\big)}{ \sqrt{\frac{L}{l_{d,f}}} }  \beta, ~~~~~~l_{d,s} < r_g < l_{d,f}, \\
       \frac{4}{3} \frac{   \ln \big(\frac{L}{r_g}\big)}{ \sqrt{\frac{L}{r_g}} }  \beta, ~~~~~~~~~~  r_g > l_{d,f} . \label{eq: lrgfsr}
\end{subnumcases}
In Fig. \ref{fig: ducismfs}, we present the estimated $D_{\mu\mu, s1,T}/D_{\mu\mu, f1,T}$ 
for CR protons in typical partially ionized interstellar phases, including the warm neutral medium (WNM), 
the cold neutral medium (CNM), the molecular cloud (MC), and the dense core in a molecular cloud (DC). 
$E_\text{CR}$ is still the CR energy corresponding to $r_g$ at $\mu = 0$. 
The parameters used and the 
ion-neutral collisional damping scales of slow and fast modes are taken from 
\citet{Xuc16}, 
as listed in Table \ref{tab: dam}. 
$D_{\mu\mu, s1,T}/D_{\mu\mu, f1,T}$ is independent of the particle energy for relatively low-energy CRs.
It increases with the CR energy for sufficiently high-energy CRs. 
In most partially ionized phases, 
$D_{\mu\mu, s1,T}/D_{\mu\mu, f1,T}$ has a small value, indicative of the dominance of fast modes in TTD scattering of CRs. 
This comes from the small value of $\beta$ and the anisotropic energy distribution of slow modes (see Section \ref{ssec: fatd}).
But in the WNM, 
it is close to unity due to the severe damping of fast modes.
In this situation, the strong damping effect makes the TTD with fast modes less efficient.

\begin{table}[h]
\renewcommand\arraystretch{1.5}
\centering
\begin{threeparttable}
\caption[]{$l_d$ of slow and fast modes in partially ionized interstellar phases}\label{tab: dam} 
  \begin{tabular}{ccccc}
     \toprule
                      &       WNM                       & CNM                                  & MC                                      & DC   \\
                      \hline
      $\beta$          &   $0.22$                   & $0.23$                               &  $0.20$                               & $0.03$      \\
     $B_0 [\mu G]$ &  $8.66$                   & $8.66$                               & $8.66$                                & $86.6$      \\
     $l_{d,s}$ [pc]  &  $3\times10^{-3}$       & $4\times10^{-2}$       & $5\times10^{-4}$                    & $1\times10^{-3}$  \\
     $l_{d,f}$  [pc]   &   $4$                          & $10^{-1}$                 & $2\times10^{-3}$                & $9\times10^{-3}$      \\
   \bottomrule
    \end{tabular}
 \end{threeparttable}
\end{table}

(2) Low-energy particles ($v_\| \ll V_\text{ph}$)

For very low-energy particles, the broadening mainly comes from the turbulent decorrelation.
The total $D_{\mu\mu,T}$ (Eq. \eqref{eq: totdut}) has an approximate expression 
(Eqs. \eqref{eq: dsfurs} and \eqref{eq: leftdd})
\begin{equation}
\begin{aligned}
   & D_{\mu\mu,T} \approx D_{\mu\mu,s2,T} + D_{\mu\mu,f2,T} \\
                         &  ~~~~~~~~~~~   =  \sqrt{2} \pi^\frac{3}{2} \frac{1}{B_0^2} v^2 (1-\mu^2)^2 \\
                  &     ~~~~  \Bigg\{   \frac{1}{4} C_s V_A^{-1} L^{-\frac{2}{3}} \ln \Big(\frac{L}{l_{\perp,\text{min},s}}\Big)    
                                   \exp \Bigg[ - \frac{V_{\text{ph},s}^2}{2 V_A^2 }\Bigg]  \\
     & ~~~~ +  \frac{2}{15}  C_f \frac{V_{Lf}^2}{V_{\text{ph},f}^3} L^{-\frac{1}{2}} 
                                   \exp{\bigg[-\frac{V_{\text{ph},f}^4 }{2 V_{Lf}^4  }\bigg]} \Bigg\} .
\end{aligned}
\end{equation}
Then we can easily compare $D_{\mu\mu,s2,T}$ with $D_{\mu\mu,f2,T}$
(Eqs. \eqref{eq: cs} and \eqref{eq: cf}), 
\begin{equation}\label{eq: ractsf}
    \frac{D_{\mu\mu,s2,T}}{D_{\mu\mu,f2,T}} = 5 \frac{\delta B_s^2}{ \delta B_f^2} \frac{V_{\text{ph},f}^3}{V_A V_{Lf}^2}
         \ln \Big(\frac{L}{l_{\perp,\text{min},s}}\Big)    
         \exp \Bigg[ \frac{V_{\text{ph},f}^4 }{2 V_{Lf}^4  }- \frac{V_{\text{ph},s}^2}{2 V_A^2 }\Bigg] ,
\end{equation}
where the relation $v_\| \ll V_\text{ph}$ is used. 
In a high-$\beta$ medium, the above expression becomes 
\begin{equation}
    \frac{D_{\mu\mu,s2,T}}{D_{\mu\mu,f2,T}} \approx \frac{5}{2}  \frac{V_{Ls}^2}{V_{Lf}^2} \beta
    \frac{c_s^3}{V_A V_{Lf}^2}
         \ln \Big(\frac{L}{l_{\perp,\text{min},s}}\Big)    
         \exp \Bigg[ \frac{c_s^4 }{2 V_{Lf}^4  }- \frac{1}{2  }\Bigg] .
\end{equation}
We further assume $V_{Ls}^2 \sim V_{Lf}^2 \sim V_A^2/3$, so that it reduces to 
\begin{equation}\label{eq: hbfsc}
    \frac{D_{\mu\mu,s2,T}}{D_{\mu\mu,f2,T}} \sim 2.7 \ln \Big(\frac{L}{l_{\perp,\text{min},s}}\Big)  \beta^\frac{5}{2}     
         \exp \bigg( \frac{9 }{8 } \beta^2 \bigg) .
\end{equation}
It clearly shows that 
$D_{\mu\mu,s2,T}$ is significantly larger than $D_{\mu\mu,f2,T}$.

In a low-$\beta$ medium, we rewrite Eq. \eqref{eq: ractsf} as 
\begin{equation}
    \frac{D_{\mu\mu,s2,T}}{D_{\mu\mu,f2,T}} = \frac{5}{2} \frac{ V_{Ls}^2}{V_{Lf}^2} \beta
    \frac{V_A^2}{ V_{Lf}^2}
         \ln \Big(\frac{L}{l_{\perp,\text{min},s}}\Big)    
         \exp \Bigg[ \frac{V_A^4 }{2 V_{Lf}^4  }- \frac{\beta}{4  }\Bigg] .
\end{equation}
We again assume $V_{Ls}^2 \sim V_{Lf}^2 \sim V_A^2/3$ and obtain 
\begin{equation}\label{eq: lbfsc}
\begin{aligned}
    \frac{D_{\mu\mu,s2,T}}{D_{\mu\mu,f2,T}}  
     & \approx \frac{15}{2}  \beta
         \ln \Big(\frac{L}{l_{\perp,\text{min},s}}\Big)    
         \exp \bigg( \frac{9}{2   }\bigg)   \\
      &   \approx 675 \ln \Big(\frac{L}{l_{\perp,\text{min},s}}\Big) \beta .
\end{aligned}
\end{equation}
Even with a small $\beta$, 
$D_{\mu\mu,s2,T}$ still tends to be larger than $D_{\mu\mu,f2,T}$.

The above comparisons show that despite the anisotropic energy distribution, slow modes are more important than fast modes in  
TTD resonance of very low-energy particles. 
The inefficiency of scattering by fast modes is attributed to the long decorrelation timescale, i.e., $\omega_\text{tur,f}^{-1}$, of fast modes.

\subsection{Resonance-broadened TTD scattering of CRs and charged grains}
\label{sec: comnum}

To illustrate the relative importance between the two resonance broadening effects for slow and fast modes  
and for different energies of particles, we next consider the TTD scattering of CR protons and 
non-relativistic charged grains in the diffuse warm ionized medium (WIM).

The typical parameters of the WIM are listed in Table \ref{tab: wim}
(see \citealt{DraL98}), 
where $n_H$ is the number density of gas, and $T$ is the temperature. 
For the interstellar turbulence, we assume that the turbulent energy is injected at a large scale by e.g., supernova explosions, 
with the turbulent speed comparable to $V_A$ and the turbulent energy equally distributed into three modes, i.e., 
$V_{Ls} \sim V_{Lf} \sim V_A/\sqrt{3}$.
Thus slow and fast modes have comparable magnetic fluctuations at $\beta \sim 1$. 
We also assume for simplicity that $\delta B_\|^2$ in $\Delta v_\|$ (Eq. \eqref{eq: volk}) is equal to the sum of $\delta B_s^2$ and $\delta B_f^2$. 
In addition, $l_d$ of Alfv\'{e}n and slow modes is the viscous damping scale 
\citep{YL04}, 
which is consistent with the inner scale of the 
observed power-law spectrum of electron density fluctuations in the WIM 
\citep{Armstrong95,CheL10}.
$l_d$ of fast modes is determined based on the calculations in 
\citet{YL04}.

\begin{table*}[!htbp]
\renewcommand\arraystretch{1.5}
\centering
\begin{threeparttable}
\caption[]{Typical parameters in the WIM}\label{tab: wim} 
  \begin{tabular}{ccccccccccc}
     \toprule
 \multirow{2}{*}{$n_H [\text{cm}^{-3}]$}   &  \multirow{2}{*}{T [K]} & \multirow{2}{*}{$B_0$ [$\mu$ G]} & \multirow{2}{*}{$c_s$ [km s$^{-1}$]} & \multirow{2}{*}{$V_A$ [km s$^{-1}$]} & \multirow{2}{*}{$\beta$} & \multirow{2}{*}{$L$ [pc]  }  & \multicolumn{3}{c}{$l_{d}$ [cm]} &    \multirow{2}{*}{ $\bigg(\frac{\langle\delta B_\|^2\rangle}{B_0^2}\bigg)^\frac{1}{4} $}   \\
 \cline{8-10}
                                                                &                                          &                                                             &                                            &                                       & &        &    Alfv\'{e}n      &  slow          &   fast        &       \\              
      \hline
                 $0.1$                                      &      $8000$                         &  $3$                                                    &  $14.8$                              &    $20.7$         &  $1.0$     &$30$  &    $5.4\times10^{5}$    &   $5.4\times10^{5}$   &  $10^{14}$   &  $0.84$   \\          
    \bottomrule
    \end{tabular}
 \end{threeparttable}
\end{table*}

(1) TeV CR protons

\begin{figure*}[htbp]
\centering   

\subfigure[TeV CRs, slow modes]{
   \includegraphics[width=8.5cm]{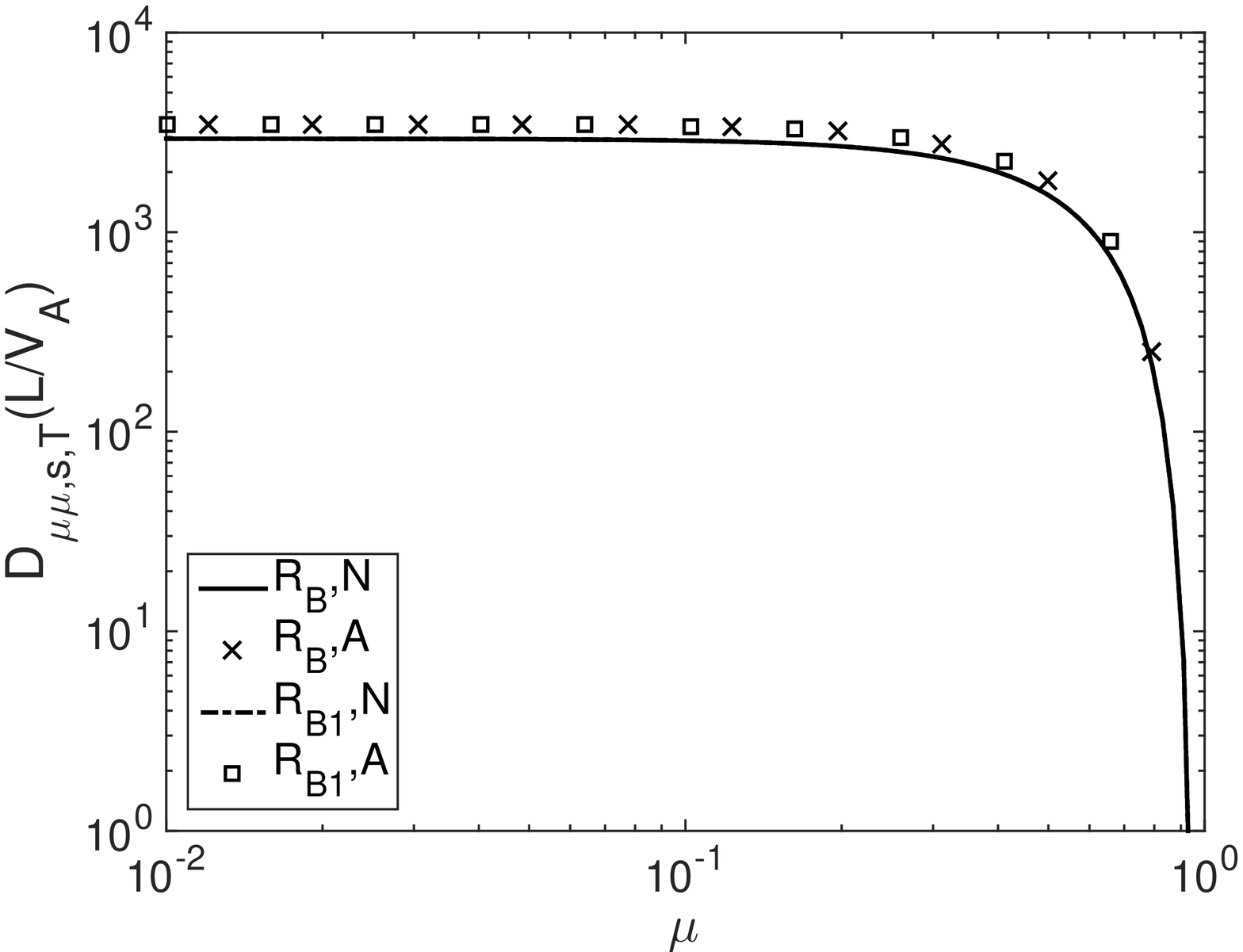}\label{fig: swimcr}}
\subfigure[TeV CRs, fast modes]{
   \includegraphics[width=8.5cm]{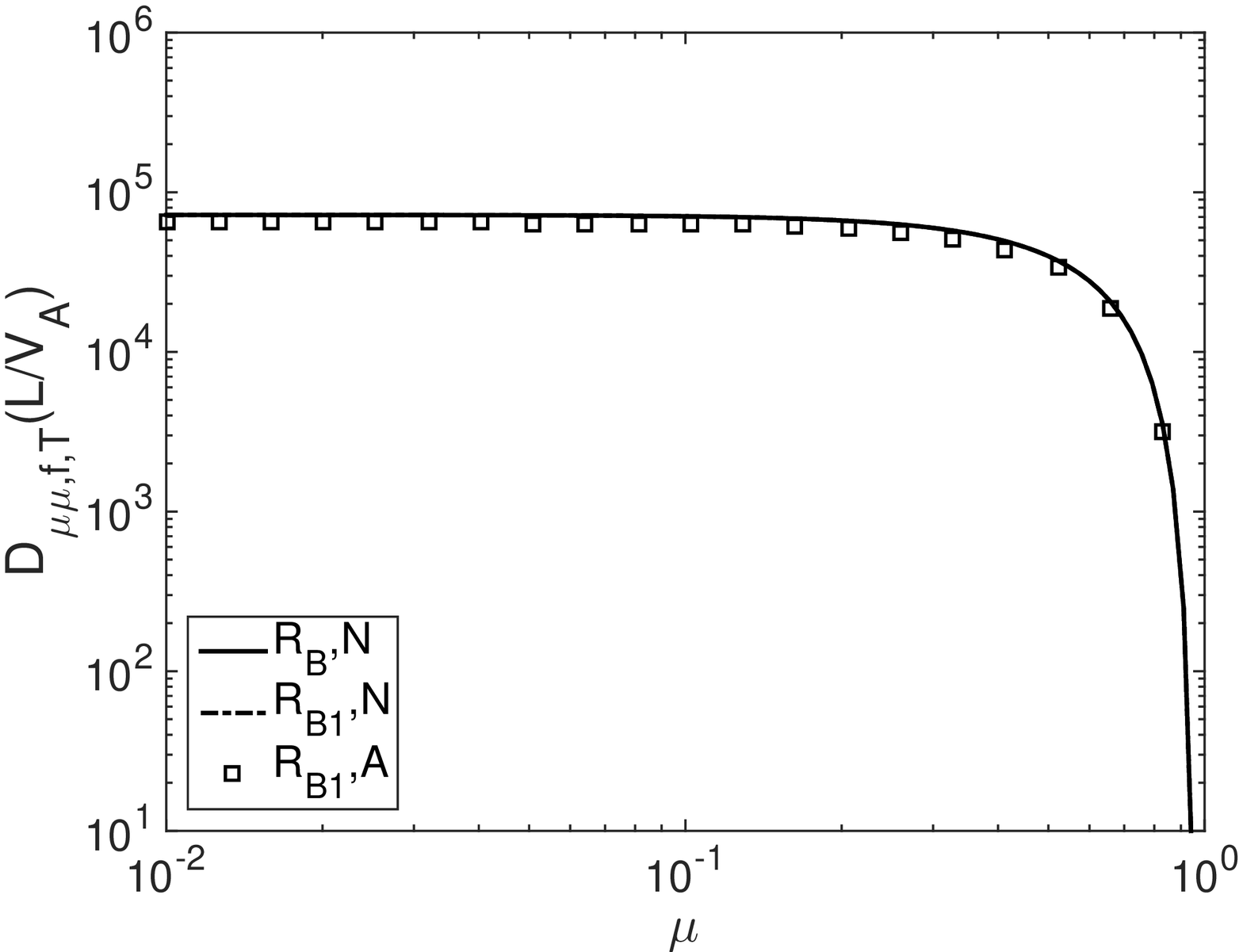}\label{fig: fwimcr}}
   
\subfigure[Charged grains with $v =0.1 V_A$, slow modes]{
   \includegraphics[width=8.5cm]{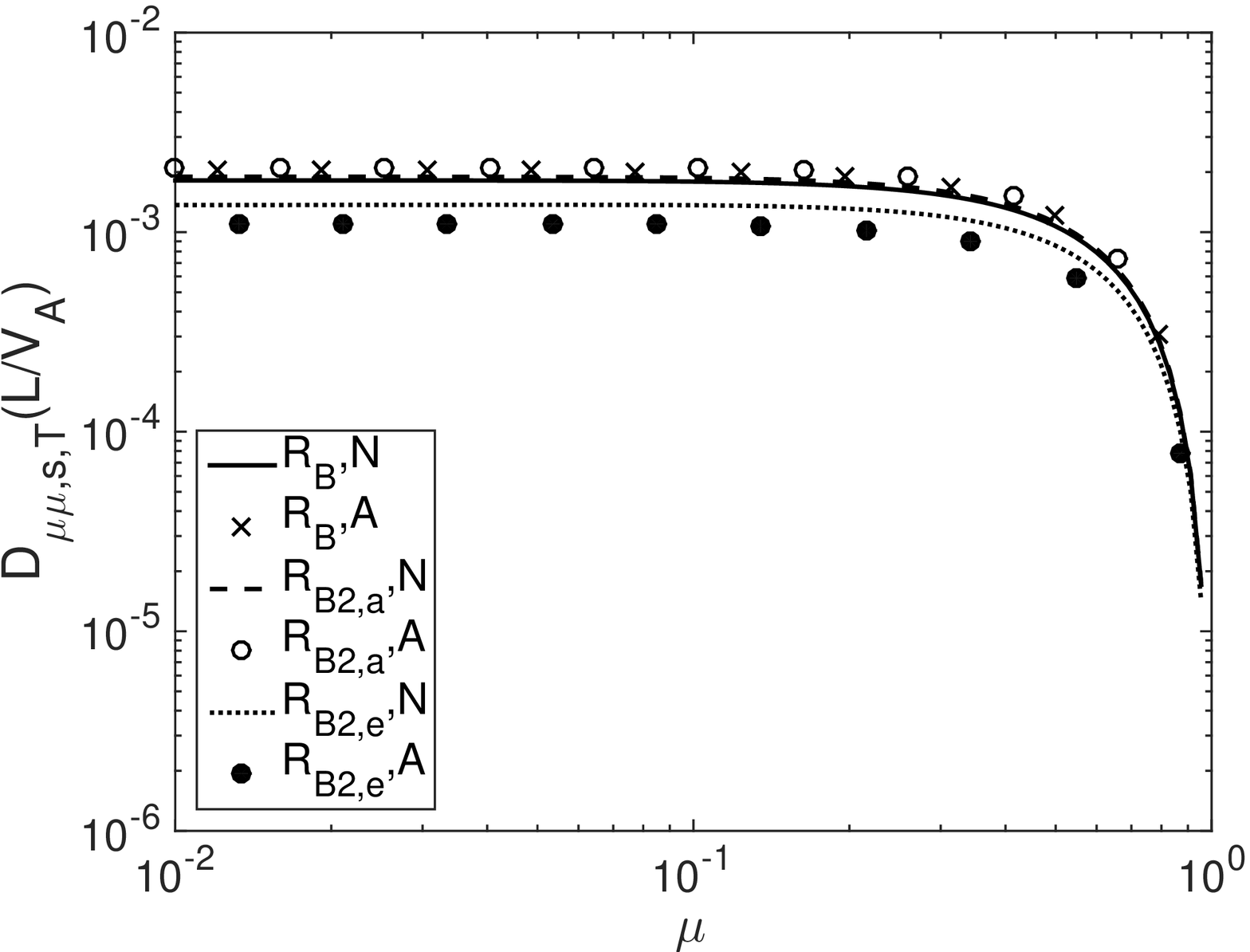}\label{fig: swimdu}}   
\subfigure[Charged grains with $v =0.1 V_A$, fast modes]{
   \includegraphics[width=8.5cm]{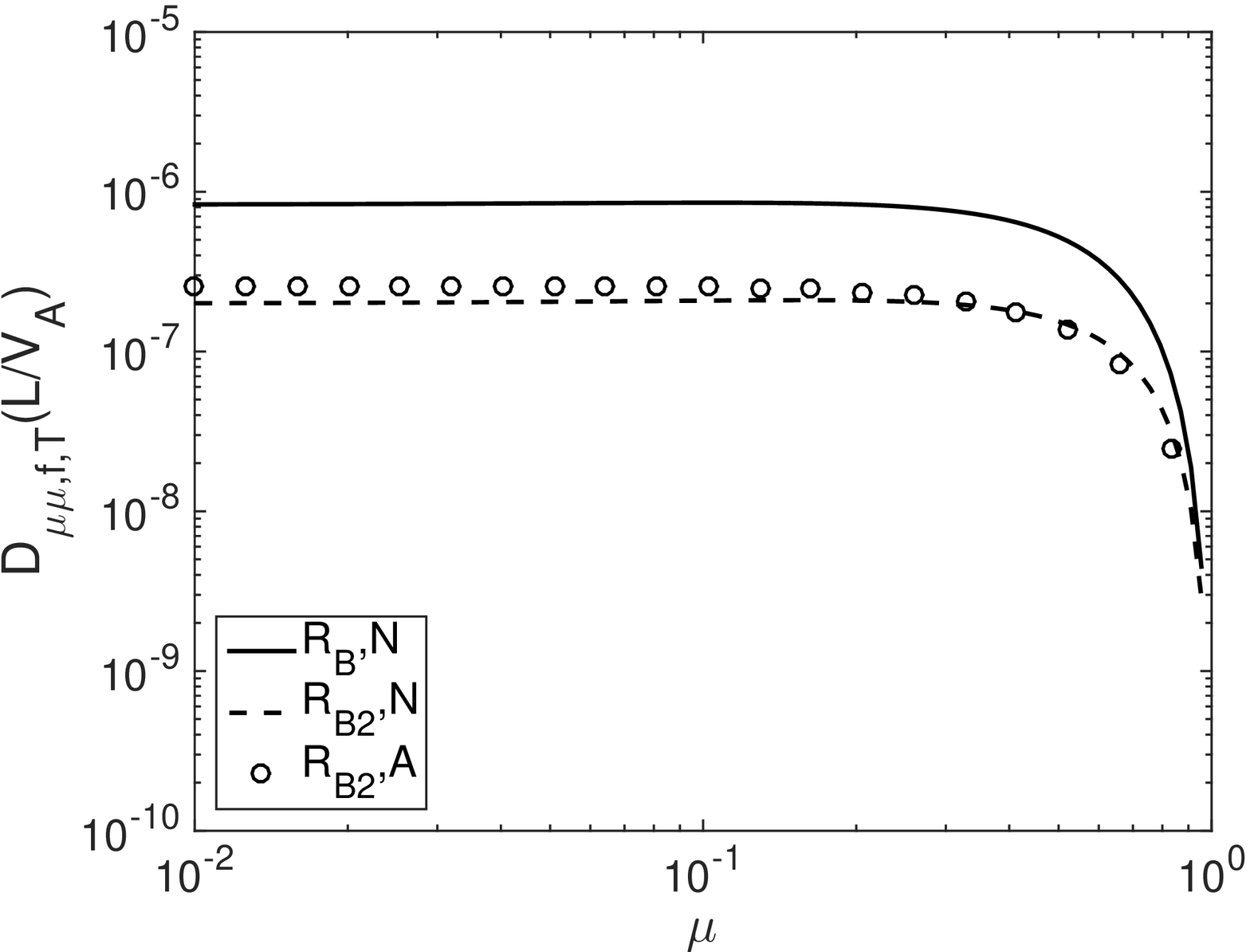}\label{fig: fwimdu}}
   
\subfigure[Charged grains with $v =1.5 V_A$, slow modes]{
   \includegraphics[width=8.5cm]{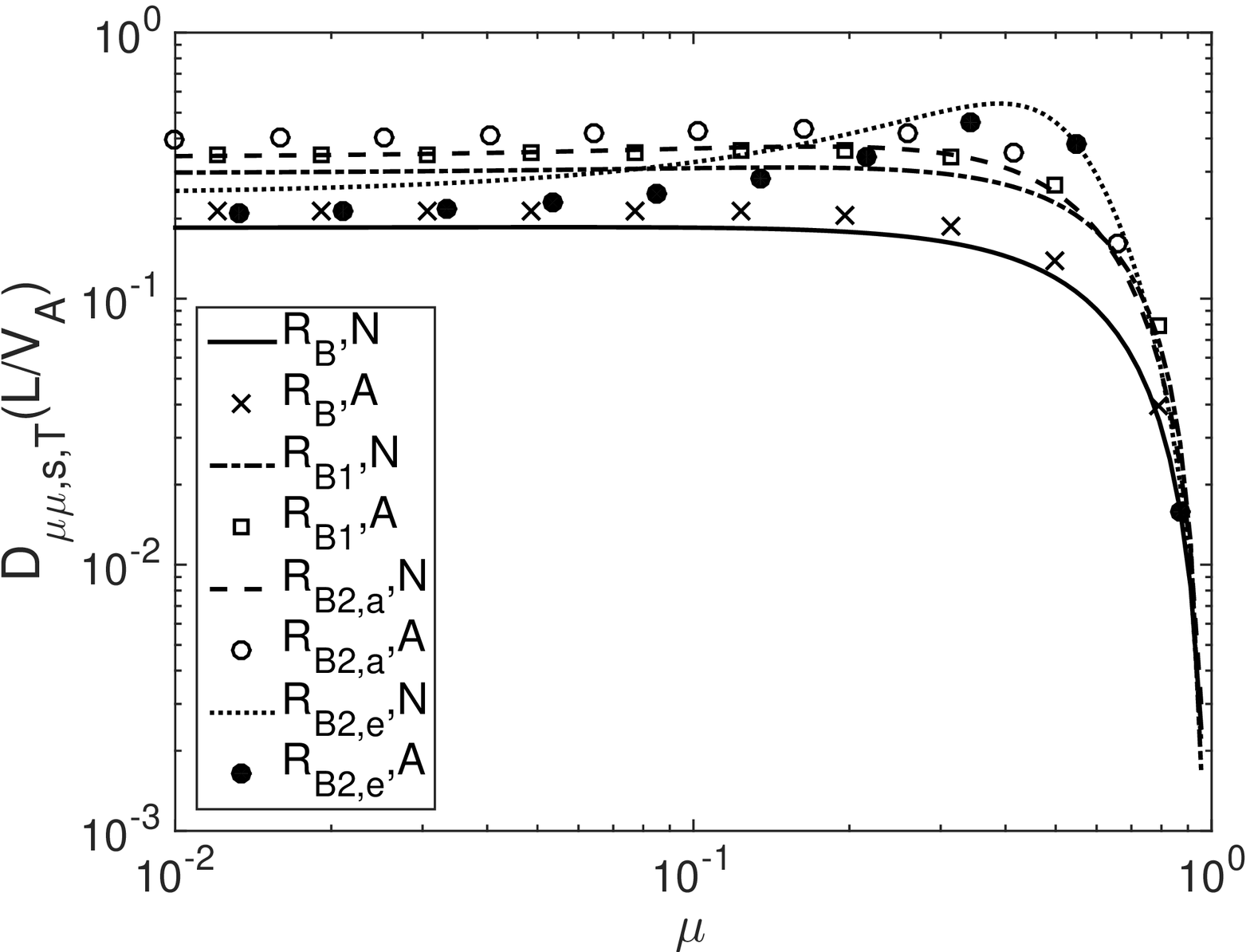}\label{fig: swimduh}}   
\subfigure[Charged grains with $v =1.5 V_A$, fast modes]{
   \includegraphics[width=8.5cm]{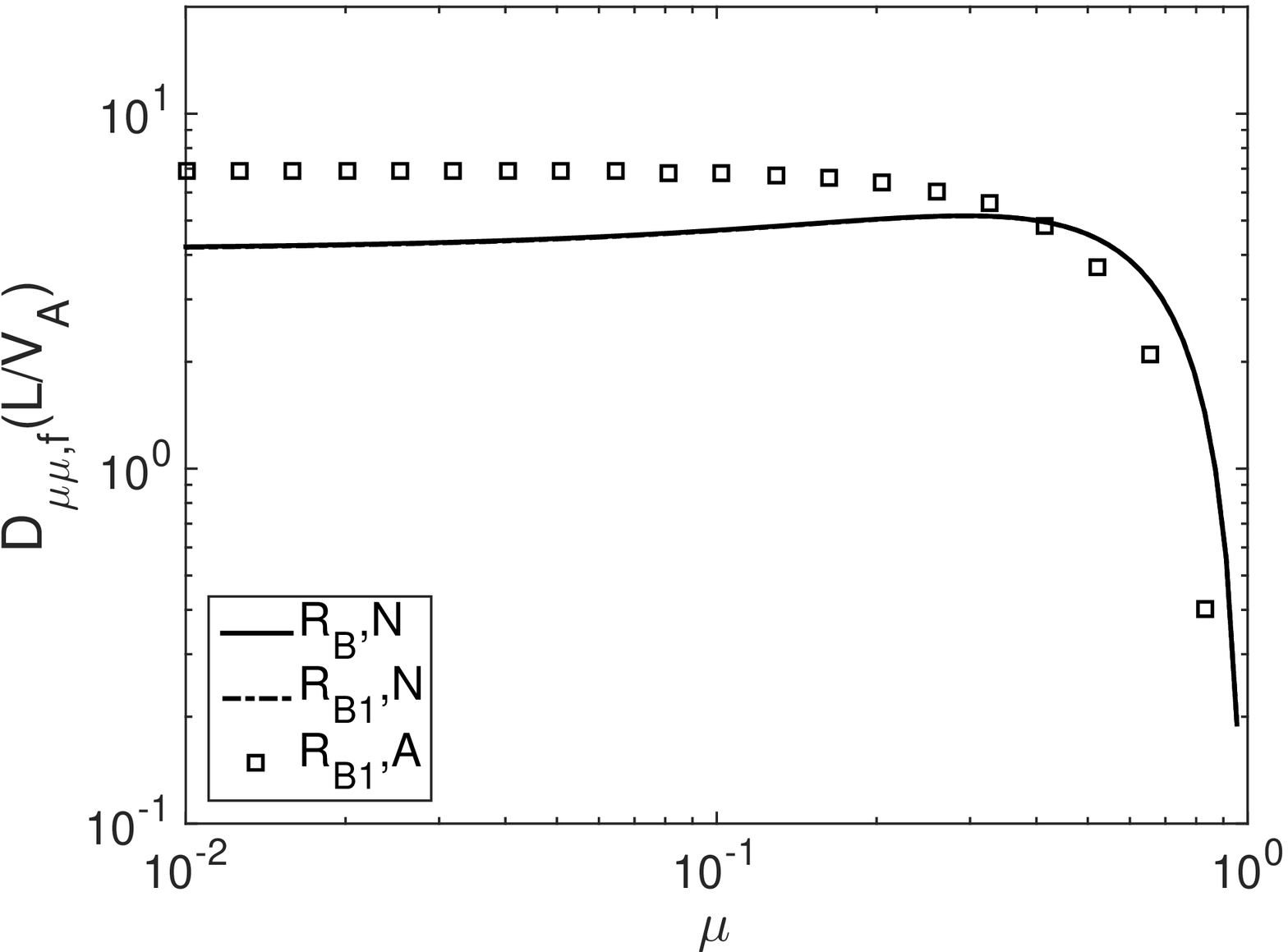}\label{fig: fwimduh}}

\caption{$D_{\mu\mu}$ for TTD with slow and fast modes in the WIM. 
``N" and ``A" represent numerical and analytical results.
For slow modes, 
$R_{B2,a}$ and $R_{B2,e}$ denote the resonance function $R_{B2}$ with the 
Alfv\'{e}n wave-crossing time and eddy-turnover time used as the turbulent decorrelation time, respectively.  }
\label{fig: ttsf}
\end{figure*}

For high-energy CRs interacting with slow modes, the condition $\Delta v_\| \gg V_A$ is satisfied for most pitch angles except for $\mu \rightarrow 1$, 
implying the dominance of the broadening due to magnetic fluctuations. 
Fig. \ref{fig: swimcr} displays 
the numerically calculated $D_{\mu\mu,s,T}$ with $R_B$ (solid line) 
and $D_{\mu\mu,s1,T}$ with $R_{B1}$ (dash-dotted line), 
which are overlapped as expected. 
Corresponding analytical approximations are Eq. \eqref{eq: sange} (crosses) and Eq. \eqref{eq: sbfurs} (squares), 
where $l_{\perp,\text{min}} = r_g$ as $r_g \gg l_{d,s}$
(e.g., $r_g$ at $\mu=0$ is $1.1\times10^{15}$ cm).

For TTD scattering by fast modes, 
Fig. \ref{fig: fwimcr} presents the overlapping $D_{\mu\mu,f,T}$ and $D_{\mu\mu,f1,T}$, 
which are numerically calculated with $R_B$ and $R_{B1}$ used, respectively. 
The analytical result for the latter case is given by Eq. \eqref{eq: fbma}, where $l_\text{min} = r_g$ is used as $r_g > l_{d,f}$.

The above results confirm that for TTD scattering of high-energy particles, the broadening due to magnetic fluctuations is dominant 
for both slow and fast modes. 
The damping of turbulence does not affect the scattering efficiency as long as $r_g$ is larger than $l_d$. 
Besides, the ratio $D_{\mu\mu, s1,T}/D_{\mu\mu, f1,T}$ can be quantified by using Eq. \eqref{eq: lrgfsr}, which provides 
$D_{\mu\mu, s1,T}/D_{\mu\mu, f1,T} \approx 0.05$, in a good agreement with the numerical result. 
It shows the dominance of fast modes in TTD scattering of TeV CRs in the WIM.

(2) Low-energy charged grains

Here we consider  
graphite grains with the density $\rho_d = 2.2$ g cm$^{-3}$, the charge $Z_d=2$, and the size $a=10^{-6}$ cm 
\citep{HLS12}.
We adopt $v = 0.1 V_A$ to exemplify the TTD scattering of very low energy particles.

In the case of slow modes, 
As $\Delta v_\| \ll V_A $, the broadening due to turbulent decorrelation is dominant.
Fig. \ref{fig: swimdu} shows that the numerically calculated $D_{\mu\mu,s,T}$ with $R_B$ (solid line) 
coincides with the numerically calculated $D_{\mu\mu,s2,T}$ with $R_{B2}$ (dashed line). 
Corresponding analytical approximations are taken from Eq. \eqref{eq: sange} (crosses) and Eq. \eqref{eq: dsfurs} (open circles), 
where $l_{\perp,\text{min}} = r_g$ as $r_g \gg l_{d,s}$ (e.g., $r_{g, \mu = 0} = 2.0\times10^{13}$ cm).
Besides, we also present the numerically (dotted line) and analytically (Eq. \eqref{eq: setcv}, filled circles) calculated $D_{\mu\mu,s2, T}$ 
using the eddy-turnover rate as $\omega_\text{tur,s}$, 
which are close to other results using the Alfv\'{e}n wave-crossing rate because of the critical balance in strong MHD turbulence.

As regards the fast modes, the numerical result of $D_{\mu\mu,f,T}$ with $R_B$ is presented in Fig. \ref{fig: fwimdu} (solid line).
The numerically (dashed line) and analytically (Eq. \eqref{eq: leftdd}, open circles) derived $D_{\mu\mu,f2,T}$ with $R_{B2}$
have a smaller value. 
It implies that  
although the main contribution comes from the broadening due to turbulent decorrelation, 
both broadening effects should be taken into account to obtain the accurate result.

The above results confirm that as a consequence of the decorrelation of turbulent magnetic fields, 
particles with $v\ll V_\text{ph}$ can also undergo the TTD resonance. 
We also see that as analyzed in Section \ref{sssec: com},
for TTD scattering of low-energy particles, slow modes are much more efficient than fast modes.

(3) Charged grains with super-Alfv\'{e}nic speeds

We again consider graphite grains but with $v=1.5 V_A$ as an example for TTD scattering of moderate-energy particles. 
As $v$ is comparable to $V_A$, 
both broadening effects are important for TTD with slow modes, and thus the general expression of $D_{\mu\mu,s,T}$ should be used. 
In Fig. \ref{fig: swimduh}, 
we see that $D_{\mu\mu,s1,T}$ with $R_{B1}$ and $D_{\mu\mu,s2,T}$ with $R_{B2}$ have similar values. 
They are both larger than the general $D_{\mu\mu,s,T}$ with $R_B$. 
The analytical expressions used here are Eq. \eqref{eq: sange} for $D_{\mu\mu,s,T}$, 
Eq. \eqref{eq: ssbf} for $D_{\mu\mu,s1,T}$, 
and Eq. \eqref{eq: dscb} for $D_{\mu\mu,s2,T}$, 
where $l_{\perp,\text{min}} = r_g$ as $r_g \gg l_{d,s}$ (e.g., $r_{g, \mu = 0} = 3.0 \times10^{14}$ cm).

For TTD with fast modes, $v \sim V_A$ indicates  
$ \Delta v_\| k_\| \sim  V_A  k_\|= \omega_\text{tur,s} > \omega_\text{tur,f}$ and the dominance of the broadening due to magnetic fluctuations. 
As expected, 
Fig. \ref{fig: fwimduh} shows that $D_{\mu\mu,f,T}$ with $R_B$ and $D_{\mu\mu,f1,T}$ with $R_{B1}$ are the same. 
The analytical approximation of $D_{\mu\mu,f1,T}$ is derived from Eq. \eqref{eq: rodfmf} in the asymptotic limit $v_\| \ll V_\text{ph}$,
\begin{equation}\label{eq: resdme}
\begin{aligned}
    D_{\mu\mu,f1,T} 
\approx & \frac{\sqrt{2}}{4}\pi^\frac{3}{2}  \frac{C_f}{B_0^2}  \Bigg(\frac{\langle\delta B_\|^2\rangle}{B_0^2}\Bigg)^{-\frac{1}{4}} \Big(l_\text{min}^{-\frac{1}{2}} - L^{-\frac{1}{2}}\Big)  v  \\
   & (1-\mu^2)^\frac{3}{2} \exp \Bigg[- \frac{  V_\text{ph}^2}{\Delta v_\|^2}\Bigg]  ,
\end{aligned}
\end{equation}
where $l_\text{min} = r_g$ as $r_g > l_{d,f}$. 
It only serves as a rough estimate of $D_{\mu\mu,f1,T}$.
By comparing with slow modes, 
we see that TTD with fast modes is more efficient in scattering moderate-energy particles in the WIM.

In addition, 
by comparing the pitch-angle diffusion coefficients at different particle energies, 
we find that as $D_{\mu\mu,T}$ depends on $v$ (see Table \ref{tab:vel}), CRs are much more efficiently scattered than charged grains by 
both slow and fast modes.

\section{ Resonance-broadened gyroresonance}

We have detailed the resonance-broadened TTD above, next 
we will briefly discuss the effect of resonance broadening on the gyroresonant pitch-angle scattering.

The linear resonance function of gyroresonance scattering is 
\begin{equation}
    R_L = \pi \delta(\omega_k - v_\| k_\| + n \Omega )  , 
\end{equation}
where $n$ is a non-zero integer. 
The strong scattering requires that $\omega_k$ is Doppler-shifted to the gyrofrequency $\Omega$ of the particle or its harmonics. 
In MHD turbulence, 
similar to the broadening of TTD resonance, the broadened resonance function for gyroresonance is 
\begin{equation}\label{eq: gegybr}
    R_B = \frac{\sqrt{2\pi}}{2 ( \Delta v_\| k_\|+ \omega_\text{tur})}
    \exp{\Bigg[-\frac{(\omega_k - v_\| k_\| +n \Omega)^2 }{2 ( \Delta v_\| k_\|+ \omega_\text{tur})^2}\Bigg]} .
\end{equation}
For high-energy particles with $v \gg V_\text{ph}$ and $\Delta v_\| k_\|  \gg \omega_\text{tur}$, the broadening induced by magnetic fluctuations is dominant. 
The above $R_B$ can be simplified, 
\begin{equation}
    R_{B1} \approx \frac{\sqrt{2\pi}}{2 \Delta v_\| k_\| }
    \exp{\Bigg[-\frac{( - v_\| k_\| + \Omega)^2 }{2 ( \Delta v_\| k_\| )^2}\Bigg]} ,
\end{equation}
with $n=1$ for the dominant resonance. 
As the magnetic perturbations on scales smaller than $r_g$ are unimportant in scattering particles, we consider $k \lesssim 1/r_g$. 
At a small $k$, the exponential term in $R_{B1}$ becomes small, and the scattering becomes inefficient. 
We see that only when $\mu$ is small, can the broadening make a significant difference compared to the linear resonance
\citep{YL08,Xuc16}.

For low-energy particles with $v\ll V_\text{ph}$ and $\Delta v_\| k_\|  \ll \omega_\text{tur}$, there is 
\begin{equation}
    R_{B2} \approx \frac{\sqrt{2\pi}}{2  \omega_\text{tur}}
    \exp{\Bigg[-\frac{(\omega_k  - \Omega)^2 }{2  \omega_\text{tur}^2}\Bigg]} ,
\end{equation}
with $n=-1$ for the dominant resonance. 
In the case of fast modes with a very small $\omega_\text{tur,f}$, the broadening effect is negligible.

We next provide the formulae of pitch-angle diffusion coefficients for gyroresonance. 

(1)  Alfv\'{e}n modes

The pitch-angle diffusion coefficient for the gyroresonance with Alfv\'{e}n modes is 
\citep{Kulsrud_Pearce,Volk:1975}
\begin{equation}
     D_{\mu\mu,A,G} = C_\mu \int d^3k x^{-2} [J_1(x)]^2 I(k) R(k),
\end{equation}
where $C_\mu$ is given in Eq. \eqref{eq: cmu}. 
The anisotropic magnetic energy spectrum is 
\citep{CLV_incomp}
\begin{equation}
     I_A(k) = C_A  k_\perp^{-\frac{10}{3}} \exp{\Bigg(-L^\frac{1}{3}\frac{k_\|}{k_\perp^\frac{2}{3}}\Bigg)},
\end{equation} 
where 
\begin{equation}
    C_A = \frac{1}{6 \pi} \delta B_A^2 L^{-\frac{1}{3}},
\end{equation}
and 
\begin{equation}
    \frac{ \delta B_A}{B_0} \sim \frac{V_{LA}}{V_A} \sim 1
\end{equation}
for strong MHD turbulence. 
Based on the critical balance, we also have 
\begin{equation}
    \omega_k =\omega_\text{tur, A} = V_A k_\| ,
\end{equation}
which is used in the broadened resonance function $R_{B2}$.

(2) Slow and fast modes 

For gyroresonance with compressive modes, the pitch-angle diffusion coefficient is 
\citep{Kulsrud_Pearce,Volk:1975}
\begin{equation}\label{eq: comgyd}
      D_{\mu\mu,sf,G} = C_\mu \int d^3k \frac{k_\|^2}{k^2} [J_1^\prime (x)]^2 I(k) R(k)  , 
\end{equation}
where 
\begin{equation}
     J_1^\prime (x) = \frac{1}{2} [J_0 (x) - J_2 (x) ].
\end{equation}
$I(k)$ is given in Eqs. \eqref{eq: slspe} and \eqref{eq: fsep} for slow and fast modes, respectively.

\begin{figure*}[htbp]
\centering   

\subfigure[TeV CRs]{
   \includegraphics[width=8.5cm]{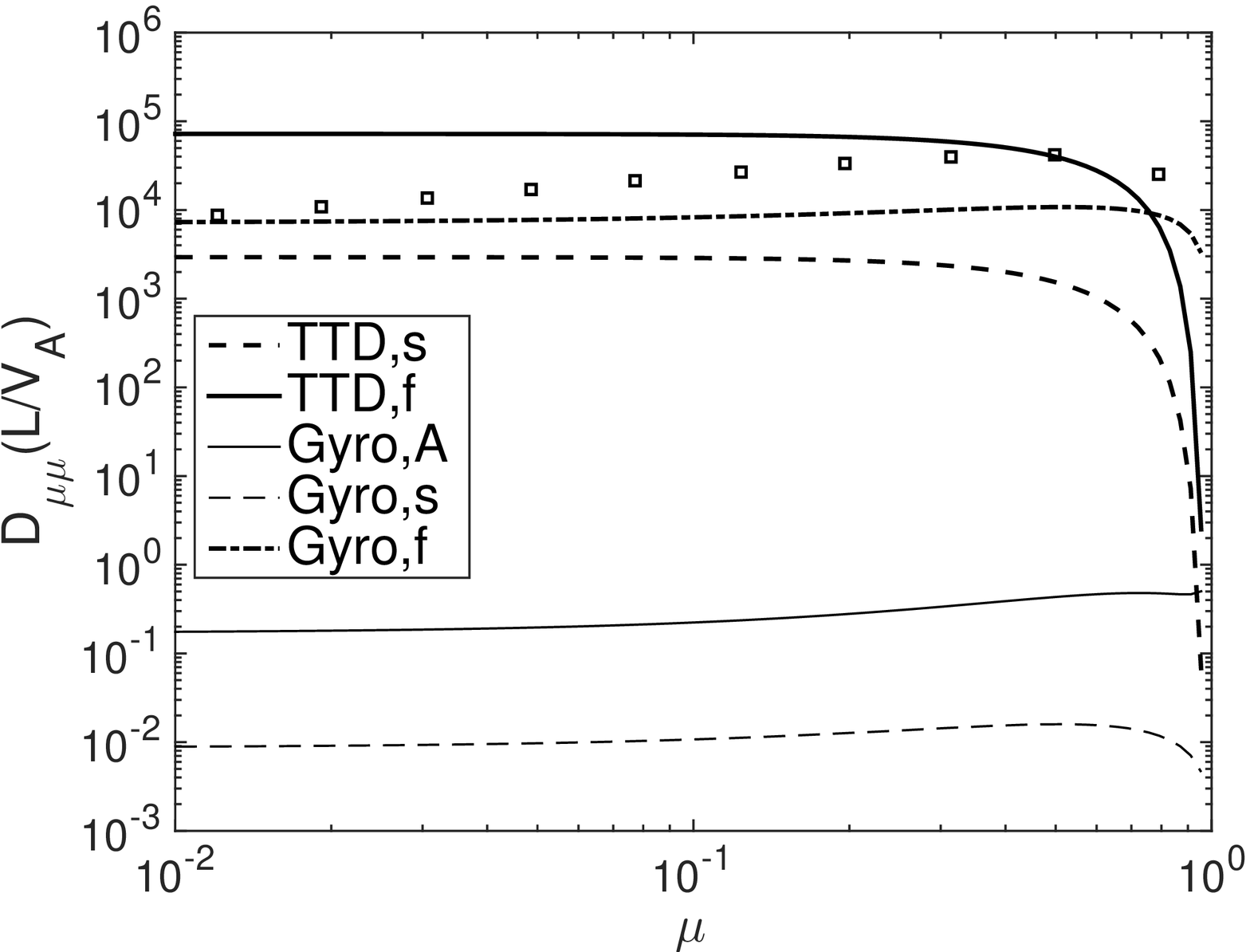}\label{fig: crall}}
\subfigure[Charged grains with $v=0.1 V_A$]{
   \includegraphics[width=8.5cm]{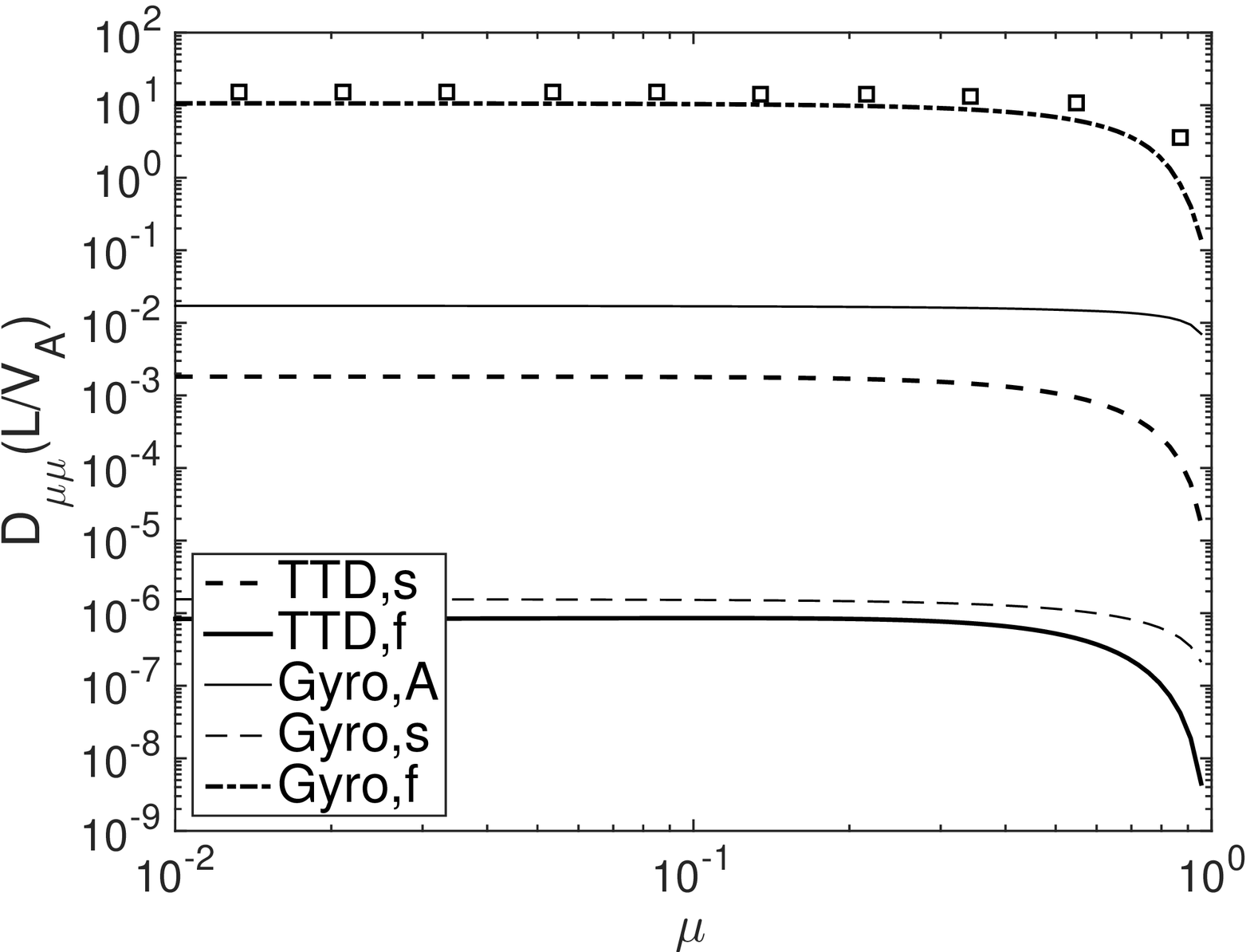}\label{fig: dustall}}
   
   \subfigure[Charged grains with $v=1.5 V_A$]{
   \includegraphics[width=8.5cm]{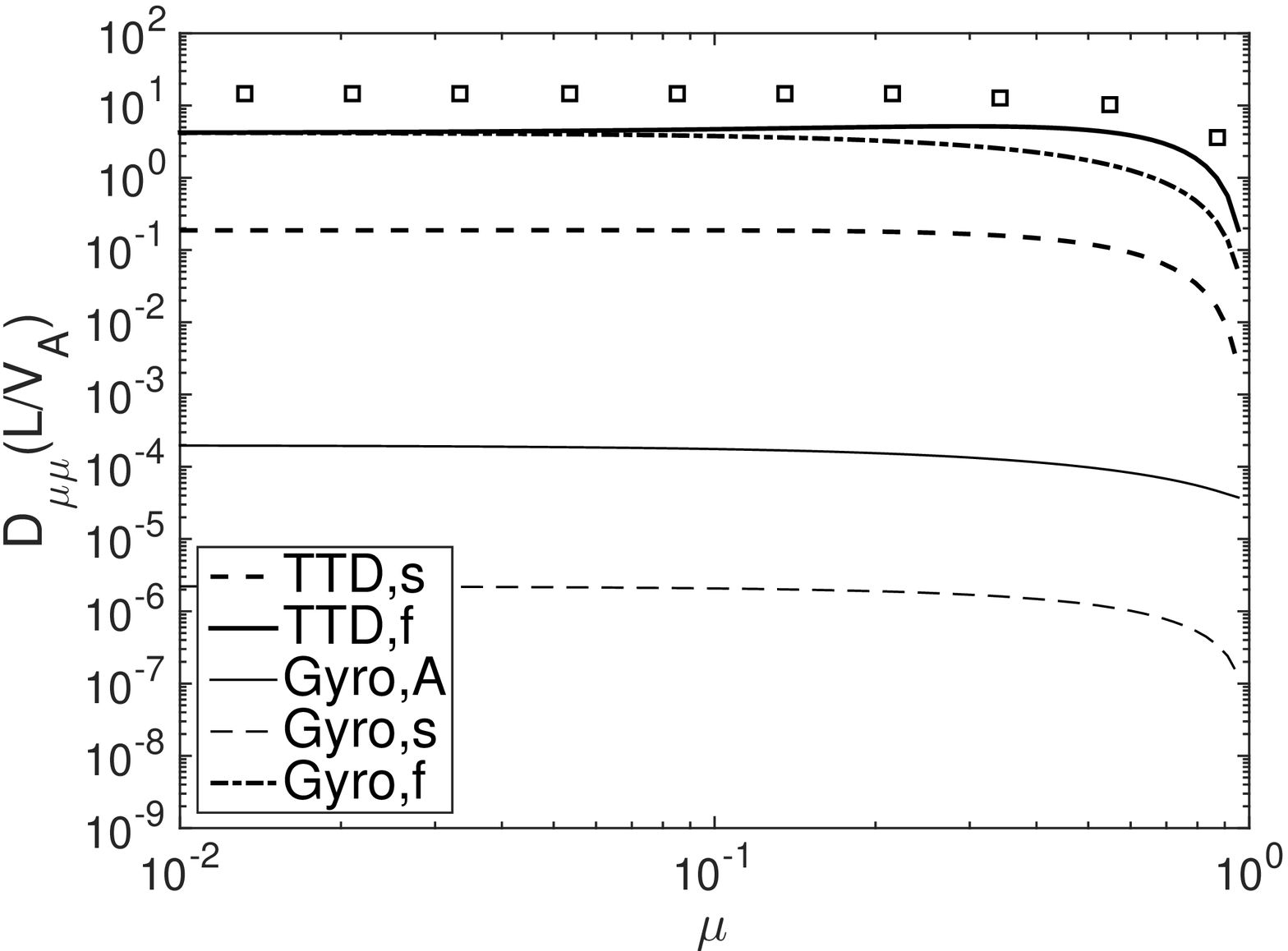}\label{fig: dustallh}}

\caption{  Numerically calculated $D_{\mu\mu}$ in the WIM. ``A,s,f" represent Alfv\'{e}n, slow, and fast modes. ``Gyro" means gyroresonance. 
The analytical approximations of gyroresonance with fast modes under the linear resonance condition are indicated by squares.
}
\label{fig: all}
\end{figure*}

For illustrative purposes, here we consider broadened gyroresonant scattering of TeV CRs and charged grains in MHD turbulence 
in the WIM, in comparison with the TTD scattering as shown in Fig. \ref{fig: ttsf}.
The parameters used are the same as those in Section \ref{sec: comnum}. 
In the case of TeV CR protons (see Fig. \ref{fig: crall}), 
due to the anisotropic scaling of Alfv\'{e}n and slow modes, the turbulent eddy at $k_\| \sim r_g^{-1}$ has $k_\perp \gg r_g^{-1}$. 
The interactions with many uncorrelated turbulent eddies within one gyro orbit 
make the gyroresonance scattering inefficient 
\citep{Chan00, YL02}.
By contrast, we see that the gyroresonance with isotropic fast modes is much more efficient, which is consistent with the earlier finding in 
\citet{YL08}.

We also obtain the analytical result from Eq. \eqref{eq: comgyd}
for gyroresonance with fast modes by using the linear resonance function 
(squares in Fig. \ref{fig: crall}, see also \citealt{Xuc16}), 
\begin{equation}
   D_{\mu\mu,fL,G} \approx \frac{2\pi^2}{7} \frac{C_f}{B_0^2} \Omega^\frac{1}{2} v^\frac{1}{2} (1-\mu^2)\mu^\frac{1}{2} .
\end{equation} 
By comparing with $D_{\mu\mu,f1,T}$ in Eq. \eqref{eq: fbma}, we find 
\begin{equation}\label{eq: crftgc}
   \frac{D_{\mu\mu,f1,T}}{D_{\mu\mu,fL,G}}
   \approx \frac{7\sqrt{2} }{8\sqrt{\pi}} \Bigg(\frac{\langle\delta B_\|^2\rangle}{B_0^2}\Bigg)^{-\frac{1}{4}} 
   \exp \Bigg[- \frac{  v_\|^2}{2\Delta v_\|^2}\Bigg]   
   \frac{        (1-\mu^2)^\frac{1}{4}   }{     \sqrt{\mu}} ,
\end{equation}
where we assume $L \gg l_\text{min} = r_g$.
It shows that TTD and gyroresonance scattering of high-energy particles with fast modes have comparable efficiencies at a moderate $\mu$, 
but TTD dominates over gyroresonance at a small $\mu$, 
and gyroresonance dominates over TTD at a large $\mu$.

In the case of low-energy graphite grains with $v=0.1 V_A$,
the resonance scale
$k_\|^{-1} (\text{or}~ k^{-1}) = (V_\text{ph} / v_\perp) r_g$
can be much larger than $r_g$.  
Fig. \ref{fig: dustall} shows that the gyroresonance with fast modes dominates the scattering. 
By using the linear resonance function, we obtain the analytical result for gyroresonance with fast modes
(squares in Fig. \ref{fig: dustall}), 
\begin{equation}\label{eq: duflg}
      D_{\mu\mu,fL,G} 
                               \approx \frac{\pi^2}{3} \frac{C_f}{B_0^2}  \Omega^{\frac{1}{2}} V_\text{ph,f}^\frac{1}{2}
                               (1 - \mu^2).
\end{equation}
It agrees well with the numerical result with the broadened resonance function, indicating the negligible role of broadening for 
gyroresonance with fast modes as mentioned above.

By comparing $D_{\mu\mu,fL,G}$ with $D_{\mu\mu,s2,T}$ in Eq. \eqref{eq: dsfurs}, we find 
\begin{equation}
\begin{aligned}
     \frac{D_{\mu\mu,s2,T}}{D_{\mu\mu,fL,G}} \approx & 2\sqrt{2} \pi^{-\frac{1}{2}}  \frac{ \delta B_s^2}{\delta B_f^2} \bigg(\frac{r_g}{L}\bigg)^{\frac{1}{2}} 
     \bigg(\frac{V_A}{V_\text{ph,f}}\bigg)^\frac{1}{2}  \bigg(  \frac{    v}{V_A}\bigg)^\frac{3}{2}     \\
                          &     \ln \Big(\frac{L}{l_{\perp,\text{min},s}}\Big)  \exp \Bigg[ - \frac{V_\text{ph,s}^2}{2 V_A^2 }\Bigg]  (1-\mu^2)^\frac{3}{4} .
\end{aligned}
\end{equation}
For low-energy particles with $r_g \ll L$ and $v \ll V_A$, 
the above ratio is small. 
As $D_{\mu\mu,f2,T}$ is even smaller than $D_{\mu\mu,s2,T}$, 
it shows that for low-energy particles in the WIM, 
gyroresonance with fast modes dominates over TTD with both slow and fast modes at all pitch angles.

In Fig. \ref{fig: dustallh}, we also present the results for graphite grains with $v=1.5 V_A$. 
We see that TTD with fast modes becomes comparably important as gyroresonance with fast modes. 
Approximately, we have (Eqs \eqref{eq: resdme} and \eqref{eq: duflg})
\begin{equation}
\begin{aligned}
     \frac{D_{\mu\mu,f1,T} }{D_{\mu\mu,fL,G} } 
\approx & \frac{3\sqrt{2}}{4\sqrt{\pi}}    \Bigg(\frac{\langle\delta B_\|^2\rangle}{B_0^2}\Bigg)^{-\frac{1}{4}}     
     \Big(\frac{v}{ V_\text{ph,f}}\Big)^\frac{1}{2}  \\
   &  \exp \Bigg[- \frac{  V_{\text{ph,f}}^2}{\Delta v_\|^2}\Bigg] (1-\mu^2)^\frac{1}{4} ,
\end{aligned}
\end{equation}
where we again assume $L \gg l_\text{min} = r_g$. The above ratio is of the order of unity.

\section{Conclusions}

The key element of studying the scattering and acceleration of particles is a physically realistic model of MHD turbulence. 
Its characteristics including 
the anisotropic scaling of Alfv\'{e}n and slow modes with respect to the {\it local} magnetic field are natural  
consequences of the nonlinear dynamics of MHD turbulence.

Due to the magnetic fluctuations and nonlinear decorrelation of turbulence, 
the broadening effects arising from MHD turbulence play an important role in enhancing the pitch-angle scattering of particles 
in MHD turbulence. 
When addressing the broadening effect induced by magnetic fluctuations, it is noteworthy that 
$\delta B_\|$ in Eq. \eqref{eq: volk} refers to 
(i) the parallel component of magnetic fields; 
(ii) the compressive component of MHD turbulence, or pseudo-Alfv\'{e}n modes in incompressible MHD turbulence; 
and (iii) the magnetic perturbation averaged over the parallel mean free path of particles when it is smaller than $L$;
which should be all considered in a more accurate treatment of the broadening. 
As regards the broadening effect caused by the turbulent decorrelation,
the long decorrelation timescale of fast modes originates from not only the spectral scaling (Eq. \eqref{eq: fsep}), but also 
the factor $v_k/V_\text{ph}$ in the cascading rate (Eq. \eqref{eq: casrfa}). 
The latter was not taken into account in 
\citet{Ly12,Lyn14}.

We mainly focused on TTD scattering and clarified that 
although the magnetic fluctuations on all scales larger than $r_g$ can contribute to TTD, 
the main contribution comes from those on $r_g$ (or $l_d$ when $l_d > r_g$). 
The pitch-angle diffusion coefficient depends on the ratio 
$L/r_g$ ($r_g> l_d$). 
Besides, it is known that gyroresonance with slow modes is inefficient because of the anisotropy of slow modes
\citep{YL02}.
Here we demonstrate that TTD with slow modes also suffers from the turbulence anisotropy, irrespective of the resonance broadening.

With the broadened resonance, both high- ($v_\| \gg V_\text{ph}$) and low-energy ($v_\| \ll V_\text{ph}$) particles can undergo 
TTD scattering with both slow and fast modes (see Table \ref{tab:vel}). 
For high-energy particles, the broadening due to magnetic fluctuations is dominant. 
The relative importance between slow and fast modes depends on their magnetic fluctuations and the plasma damping (Eq. \eqref{eq: genradfs}). 
Despite the anisotropy of slow modes, in the following scenarios:
(i) in a high-$\beta$ medium, e.g., the intracluster medium;
(ii) when fast modes are severely damped in, e.g., partially ionized WNM (Fig. \ref{fig: ducismfs});
(iii) when the particle energy is sufficiently high (Fig. \ref{fig: dube}),
slow modes can be comparably or even more efficient than fast modes in TTD scattering.
For very low-energy particles, the broadening is determined by the turbulent decorrelation,
and slow modes always dominate over fast modes in TTD.
For moderate-energy particles with $v_\| \sim V_\text{ph}$, 
the two broadening effects can be both important for slow modes, 
while the broadening due to magnetic fluctuations becomes dominant for fast modes.
Important implications concerning e.g., dynamics of charged grains,
heating of particles, and TTD damping of MHD turbulence will be investigated in our future work.

Compared with TTD by fast modes, 
the resonance broadening has an insignificant effect on gyroresonance with fast modes. 
For high-energy particles, the relative importance between TTD and gyroresonance with fast modes depends on $\mu$ (Eq. \eqref{eq: crftgc}).
For very low-energy particles, gyroresonance with fast modes tends to be dominant in scattering at all pitch angles. 
These findings agree with earlier studies, e.g. 
\citet{YL03,YL08}.
For moderate-energy particles, TTD and gyroresonance with fast modes are comparably important.
\\
\\

S.X. acknowledges the support for Program number HST-HF2-51400.001-A provided by NASA through a grant from the Space Telescope Science Institute, which is operated by the Association of Universities for Research in Astronomy, Incorporated, under NASA contract NAS5-26555.
A.L. acknowledges the support from grant
NSF DMS 1622353.

\appendix
\section{$D_{\mu\mu,s2,T}$ calculated using the turbulent eddy-turnover rate as $\omega_\text{tur,s}$}
\label{aa1: tur}

Following the turbulent energy cascade, $\omega_\text{tur,s}$ increases toward smaller scales.  
It can be formulated as 
\begin{equation}\label{eq: carsl}
    \omega_\text{tur,s} = k_\perp u_k = V_{A} L^{-\frac{1}{3}} k_\perp^\frac{2}{3}
\end{equation}
according to the Kolmogorov scaling in the direction perpendicular to the {\it local} magnetic field. 
$V_A$ is the turbulent speed of strong Alfv\'{e}nic turbulence at $L$. 
Alfv\'{e}n modes govern the energy cascade of slow modes. 
With the above expression of $\omega_\text{tur,s}$ used, we have 
\begin{equation}\label{eq: duus2}
 D_{\mu\mu,s2,T} = \frac{\sqrt{2\pi}}{8}C_\mu C_s  \frac{v_\perp^2}{\Omega^2} V_{A}^{-1} L^\frac{1}{3}   
  \int d^3k  \frac{k_\|^2}{k_\|^2 + k_\perp^2} k_\perp^{-2} \exp \Bigg[-L^\frac{1}{3} \frac{k_\|}{k_\perp^\frac{2}{3}} - 
\frac{(V_\text{ph}-v_\|)^2k_\|^2}{2 V_{A}^2 L^{-\frac{2}{3}}k_\perp^\frac{4}{3}}\Bigg] . 
\end{equation}
When $v_\| \sim V_\text{ph}$, we obtain the approximate expression 
\begin{equation}\label{eq: convd}
   D_{\mu\mu,s2,T} \approx  \frac{\sqrt{2}}{2} \pi^\frac{3}{2} \frac{C_s}{B_0^2} V_{A}^{-1} L^{-\frac{2}{3}}  \ln \Big(\frac{L}{l_{\perp,\text{min}}}\Big) v^2 
    (1-\mu^2)^2 \bigg[1- \frac{6(V_\text{ph}-v_\|)^2 }{ V_{A}^2 } \bigg] .
\end{equation}
We notice that for particles with $v_\| \sim V_\text{ph}$, the expression of $D_{\mu\mu,s2,T}$
using $\omega_\text{tur,s} = k_\| V_A$  in Eq. \eqref{eq: dscb} can be approximated by 
\begin{equation}
    D_{\mu\mu,s2,T} \approx \frac{\sqrt{2}}{4} \pi^\frac{3}{2} \frac{C_s}{B_0^2} V_A^{-1} L^{-\frac{2}{3}} \ln \Big(\frac{L}{l_{\perp,\text{min}}}\Big) v^2   
    (1-\mu^2)^2   \bigg( 1- \frac{(V_\text{ph}-v_\|)^2}{2 V_A^2 }\bigg),
\end{equation}
which has a form similar to Eq. \eqref{eq: convd}.

When $v_\|$ is significantly away from $V_\text{ph}$, i.e., $v_\| \ll V_\text{ph}$,
we drop the first term of the exponential function in Eq. \eqref{eq: duus2} and find 
\begin{equation}
     D_{\mu\mu,s2,T} \approx  \frac{\pi^2}{4}  \frac{C_s}{B_0^2} V_{A}^2 L^{-\frac{2}{3}} \ln \Big(\frac{L}{l_{\perp,\text{min}}}\Big) v^2  
      (1-\mu^2)^2 |V_\text{ph} - v_\| |^{-3} .
\end{equation}
A similar scaling relation between $D_{\mu\mu,s2,T}$ and $v_\|$ was earlier found by  
\citet{Ly12}.
To make it converge with Eq. \eqref{eq: convd} at $v_\| = V_\text{ph}$, we modify the above expression to 
\begin{equation}\label{eq: setcv}
   D_{\mu\mu,s2,T} \approx  \frac{\sqrt{2}}{2} \pi^\frac{3}{2} \frac{C_s}{B_0^2} V_{A}^{-1} L^{-\frac{2}{3}}  \ln \Big(\frac{L}{l_{\perp,\text{min}}}\Big) v^2 
    (1-\mu^2)^2 \bigg(1+ \frac{|V_\text{ph} - v_\| |}{V_{A}}\bigg)^{-3} .
\end{equation}

\section{$D_{\mu\mu,f1,T}$ calculated with a different power-law spectrum}
\label{asec: fs}

We assume the power spectrum of fast modes is 
\begin{equation}\label{eq: fasls}
     I_f(k) = C_f k^{-\frac{11}{3}},
\end{equation}
with the normalization constant
\begin{equation}
    C_f = \frac{1}{12 \pi} \delta B_f^2 L^{-\frac{2}{3}}. 
\end{equation}
It has the same power-law index as that of slow modes in terms of $k_\perp$ (Eq. \eqref{eq: slspe}), 
but has an isotropic energy distribution. 
Combining Eqs. \eqref{eq:gel}, \eqref{eq: vbmf}, and \eqref{eq: fasls}, we approximately have
\begin{equation}
    D_{\mu\mu,f1,T}^* =  \frac{3\sqrt{2}}{8} \pi^\frac{3}{2}   \frac{C_f }{B_0^2}     \Bigg(\frac{\langle\delta B_\|^2\rangle}{B_0^2}\Bigg)^{-\frac{1}{4}} \Big(l_\text{min}^{-\frac{1}{3}} - L^{-\frac{1}{3}}\Big) v (1 - \mu^2)^\frac{3}{2} \exp{\Bigg[-\frac{   v_\|^2 }{2  \Delta v_\|^2 }\Bigg]} 
\end{equation}
for high-energy particles with $v_\| \gg V_\text{ph}$.

By comparing with $D_{\mu\mu,s1,T}$ in Eq. \eqref{eq: ssbf}, we find that $D_{\mu\mu,f1,T}^*$ depends on 
$(L/l_\text{min})^{1/3}$, while $D_{\mu\mu,s1,T}$ depends on $\ln (L/l_{\perp,\text{min}})$. 
In the case of $L \gg l_\text{min}$, fast modes with the same spectral index as that of slow modes are still much more 
effective in TTD scattering of energetic particles. 

By comparing with $D_{\mu\mu,f1,T}$ in Eq. \eqref{eq: fbma}, which depends on $(L/l_\text{min})^{1/2}$,
we see that because of the steeper spectrum used here, $D_{\mu\mu,f1,T}^*$ is smaller than $D_{\mu\mu,f1,T}$ 
due to the relatively small magnetic perturbations on small scales.

\bibliographystyle{apj.bst}
\bibliography{xu}

\begin{thebibliography}{42}
\expandafter\ifx\csname natexlab\endcsname\relax\def\natexlab#1{#1}\fi

\bibitem[{{Armstrong} {et~al.}(1995){Armstrong}, {Rickett}, \&
  {Spangler}}]{Armstrong95}
{Armstrong}, J.~W., {Rickett}, B.~J., \& {Spangler}, S.~R. 1995, \apj, 443, 209

\bibitem[{{Brunetti} \& {Lazarian}(2011)}]{BruLaz11}
{Brunetti}, G., \& {Lazarian}, A. 2011, \mnras, 412, 817

\bibitem[{{Chandran}(2000)}]{Chan00}
{Chandran}, B.~D.~G. 2000, Physical Review Letters, 85, 4656

\bibitem[{{Chepurnov} \& {Lazarian}(2010)}]{CheL10}
{Chepurnov}, A., \& {Lazarian}, A. 2010, \apj, 710, 853

\bibitem[{{Cho} \& {Lazarian}(2002)}]{CL02_PRL}
{Cho}, J., \& {Lazarian}, A. 2002, Physical Review Letters, 88, 245001

\bibitem[{{Cho} \& {Lazarian}(2003)}]{CL03}
---. 2003, \mnras, 345, 325

\bibitem[{{Cho} \& {Lazarian}(2005)}]{CL05}
---. 2005, Theoretical and Computational Fluid Dynamics, 19, 127

\bibitem[{{Cho} {et~al.}(2002){Cho}, {Lazarian}, \& {Vishniac}}]{CLV_incomp}
{Cho}, J., {Lazarian}, A., \& {Vishniac}, E.~T. 2002, \apj, 564, 291

\bibitem[{{Cho} \& {Vishniac}(2000)}]{CV00}
{Cho}, J., \& {Vishniac}, E.~T. 2000, \apj, 539, 273

\bibitem[{{Draine} \& {Lazarian}(1998)}]{DraL98}
{Draine}, B.~T., \& {Lazarian}, A. 1998, \apjl, 494, L19

\bibitem[{{Giacalone} \& {Jokipii}(1999)}]{Giacalone_Jok1999}
{Giacalone}, J., \& {Jokipii}, J.~R. 1999, \apj, 520, 204

\bibitem[{{Goldreich} \& {Sridhar}(1995)}]{GS95}
{Goldreich}, P., \& {Sridhar}, S. 1995, \apj, 438, 763

\bibitem[{{Goldstein}(1976)}]{Goldstein:1976}
{Goldstein}, M.~L. 1976, \apj, 204, 900

\bibitem[{{Hoang} {et~al.}(2012){Hoang}, {Lazarian}, \& {Schlickeiser}}]{HLS12}
{Hoang}, T., {Lazarian}, A., \& {Schlickeiser}, R. 2012, \apj, 747, 54

\bibitem[{{Jokipii}(1966)}]{Jokipii1966}
{Jokipii}, J.~R. 1966, \apj, 146, 480

\bibitem[{{Jones} {et~al.}(1973){Jones}, {Birmingham}, \&
  {Kaiser}}]{Jones:1973}
{Jones}, F.~C., {Birmingham}, T.~J., \& {Kaiser}, T.~B. 1973, \apjl, 180, L139

\bibitem[{{K{\'o}ta} \& {Jokipii}(2000)}]{Kota_Jok2000}
{K{\'o}ta}, J., \& {Jokipii}, J.~R. 2000, \apj, 531, 1067

\bibitem[{{Kowal} {et~al.}(2012){Kowal}, {Lazarian}, {Vishniac}, \&
  {Otmianowska-Mazur}}]{KL12}
{Kowal}, G., {Lazarian}, A., {Vishniac}, E.~T., \& {Otmianowska-Mazur}, K.
  2012, Nonlinear Processes in Geophysics, 19, 297

\bibitem[{{Kulsrud} \& {Pearce}(1969)}]{Kulsrud_Pearce}
{Kulsrud}, R., \& {Pearce}, W.~P. 1969, \apj, 156, 445

\bibitem[{{Lazarian} \& {Vishniac}(1999)}]{LV99}
{Lazarian}, A., \& {Vishniac}, E.~T. 1999, \apj, 517, 700

\bibitem[{{Lazarian} \& {Yan}(2002)}]{LY02}
{Lazarian}, A., \& {Yan}, H. 2002, \apjl, 566, L105

\bibitem[{{Lithwick} \& {Goldreich}(2001)}]{LG01}
{Lithwick}, Y., \& {Goldreich}, P. 2001, \apj, 562, 279

\bibitem[{Longair(1997)}]{Longairbook}
Longair, M.~S. 1997, High energy astrophysics. Volume 2: Stars, the galaxy and
  the interstellar medium (Cambridge U Press, 1994)

\bibitem[{{Lynn} {et~al.}(2012){Lynn}, {Parrish}, {Quataert}, \&
  {Chandran}}]{Ly12}
{Lynn}, J.~W., {Parrish}, I.~J., {Quataert}, E., \& {Chandran}, B.~D.~G. 2012,
  \apj, 758, 78

\bibitem[{{Lynn} {et~al.}(2014){Lynn}, {Quataert}, {Chandran}, \&
  {Parrish}}]{Lyn14}
{Lynn}, J.~W., {Quataert}, E., {Chandran}, B.~D.~G., \& {Parrish}, I.~J. 2014,
  \apj, 791, 71

\bibitem[{{Maron} \& {Goldreich}(2001)}]{MG01}
{Maron}, J., \& {Goldreich}, P. 2001, \apj, 554, 1175

\bibitem[{{Matthaeus} {et~al.}(1990){Matthaeus}, {Goldstein}, \&
  {Roberts}}]{Mat90}
{Matthaeus}, W.~H., {Goldstein}, M.~L., \& {Roberts}, D.~A. 1990, \jgr, 95,
  20673

\bibitem[{{Palmer}(1982)}]{Palmer:1982}
{Palmer}, I.~D. 1982, Reviews of Geophysics and Space Physics, 20, 335

\bibitem[{{Pines} \& {Bohm}(1952)}]{Pin52}
{Pines}, D., \& {Bohm}, D. 1952, Physical Review, 85, 338

\bibitem[{{Schlickeiser}(1994)}]{Sch94}
{Schlickeiser}, R. 1994, \apjs, 90, 929

\bibitem[{{Schlickeiser}(2002)}]{Schlickeiser02}
---. 2002, {Cosmic Ray Astrophysics}, ed. R.~{Schlickeiser}

\bibitem[{{Shalchi} {et~al.}(2004){Shalchi}, {Bieber}, {Matthaeus}, \&
  {Qin}}]{Sha04}
{Shalchi}, A., {Bieber}, J.~W., {Matthaeus}, W.~H., \& {Qin}, G. 2004, \apj,
  616, 617

\bibitem[{{Summers} \& {Ma}(2000)}]{Sum00}
{Summers}, D., \& {Ma}, C.-y. 2000, \jgr, 105, 15887

\bibitem[{{Voelk}(1975)}]{Volk:1975}
{Voelk}, H.~J. 1975, Reviews of Geophysics and Space Physics, 13, 547

\bibitem[{{V{\"o}lk}(1973)}]{Volk:1973}
{V{\"o}lk}, H.~J. 1973, \apss, 25, 471

\bibitem[{{Wanner} \& {Wibberenz}(1993)}]{Wan93}
{Wanner}, W., \& {Wibberenz}, G. 1993, \jgr, 98, 3513

\bibitem[{{Xu} {et~al.}(2016){Xu}, {Yan}, \& {Lazarian}}]{Xuc16}
{Xu}, S., {Yan}, H., \& {Lazarian}, A. 2016, \apj, 826, 166

\bibitem[{{Yan} \& {Lazarian}(2002)}]{YL02}
{Yan}, H., \& {Lazarian}, A. 2002, Physical Review Letters, 89, B1102+

\bibitem[{{Yan} \& {Lazarian}(2003)}]{YL03}
---. 2003, \apjl, 592, L33

\bibitem[{{Yan} \& {Lazarian}(2004)}]{YL04}
---. 2004, \apj, 614, 757

\bibitem[{{Yan} \& {Lazarian}(2008)}]{YL08}
---. 2008, \apj, 673, 942

\bibitem[{{Yan} {et~al.}(2004){Yan}, {Lazarian}, \& {Draine}}]{YLD04}
{Yan}, H., {Lazarian}, A., \& {Draine}, B.~T. 2004, \apj, 616, 895

\end{thebibliography}

\end{document}